\documentclass[useAMS,usenatbib,twocolumn]{mn2e}
\usepackage[]{graphicx}
\usepackage{url}
\usepackage{amssymb}
\include{aas_abbrev}

\title[X-ray absorption evolution in Gamma-Ray Bursts]
{X-ray absorption evolution in Gamma-Ray Bursts: intergalactic medium or evolutionary signature of their host galaxies?}

\author[Starling et al.]
{R.L.C. Starling$^1$\thanks{E-mail: rlcs1@le.ac.uk}, R. Willingale$^1$, N.R. Tanvir$^1$, A.E. Scott$^1$, K. Wiersema$^1$,\and P.T. O'Brien$^1$, A.J. Levan$^2$ and G.C. Stewart$^1$
\\
$^1$Department of Physics and Astronomy, University of Leicester, LE1 7RH, UK\\
$^2$Department of Physics, University of Warwick, Coventry, CV4 7AL, UK}

\begin{document}
\date{Accepted  Received ; in original form }

\pagerange{\pageref{firstpage}--\pageref{lastpage}} \pubyear{2013}

\maketitle

\label{firstpage}

\begin{abstract}
The intrinsic X-ray emission of Gamma-Ray Bursts (GRBs) is often found to be absorbed over and above the column density through our own galaxy. 
The extra component is usually assumed to be due to absorbing gas lying within the host galaxy of the GRB itself. 
There is an apparent correlation between the equivalent column density of hydrogen, $N_{\rm H,intrinsic}$ (assuming it to be at the GRB redshift), and redshift, $z$, with the few $z>6$ GRBs showing the greatest intrinsic column densities.
 We investigate the $N_{\rm H,intrinsic} - z$ relation using a large sample of {\em Swift} GRBs, as well as active galactic nuclei (AGN) and quasar samples, paying particular attention to the spectral energy distributions of the two highest redshift GRBs. Various possible sample biases and systematics that might produce such a correlation are considered, and we conclude that the correlation is very likely to be real. This may indicate either an evolutionary effect in the host galaxy properties,
or a contribution from gas along the line-of-sight, in the diffuse intergalactic medium (IGM) or intervening absorbing clouds.
Employing a more realistic model for IGM absorption than in previous works, we find that this may explain much of the observed opacity at $z \gtrsim 3$ providing it is not too hot, likely between 10$^5$\,K and 10$^{6.5}$\,K, and moderately metal enriched, $Z\sim0.2$~Z$_{\odot}$. This material could therefore constitute the Warm Hot Intergalactic Medium. However, a comparable level of absorption is also expected from the cumulative effect of intervening cold gas clouds, and given current uncertainties it is not possible to say which, if either, dominates.
At lower redshifts, we conclude that gas in the host galaxies must be the dominant contributor to the observed X-ray absorption.
\end{abstract}


\section{Introduction}

\label{s1}
Gamma-Ray Burst (GRB) afterglows are thought to arise as the GRB jet 
impacts and shocks the surrounding medium (e.g. M\'esz\'aros \& Rees 1992,1997; Sari, Piran \& Narayan 1998), producing synchrotron radiation at all wavelengths from X-ray through to radio. 
X-ray emission (0.3--10\,keV) is 
routinely monitored by the {\it Swift} satellite (Gehrels et al. 2004; Burrows et al. 2005),  
and usually displays a spectral shape that is well modelled by a power-law continuum with absorption at soft (low) energies.
Some photoelectric absorption is expected due to gas in the Milky Way along the line of sight,
and at Solar abundances this effect is dominated by absorption due to metals.  
By convention, though, the absorption is parameterised by the equivalent column density of
hydrogen, assumed to be at zero redshift and Solar metallicity (Z$_{\odot}$), unless otherwise stated.

However, most
{\em Swift} GRBs show evidence for significant absorption in excess of the Galactic foreground.
This excess has usually been assumed to be caused by gas in the host galaxy of the GRB.
Since the majority of GRBs are produced by the deaths of massive stars, which are frequently found in dense
star-forming regions, a high column of gas along the line of sight would not be surprising.
Furthermore, optical spectra, when available, usually also show the signature of significant
gas columns, particularly in the form of strong (frequently damped) Lyman-$\alpha$ absorption \citep[e.g.,][]{Fynbo}.
In fact, the inferred column from X-rays is often in excess of that seen in the optical. This is presumed to be due to hydrogen very close to the burst (i.e. a few to tens of parsecs)
being ionised, either
by the GRB itself or the ambient radiation field, not contributing to the optical absorption,
whereas the X-rays are attenuated by interactions with more strongly bound electrons in 
heavier atoms and ions.

A key technical point, however, is that the X-ray absorbing effect of a given column of gas
is expected to decrease with increasing redshift, since for a fixed observed energy
passband one is observing increasingly energetic photons in the rest-frame, for
which the absorption cross section is reduced.
Specifically, the mapping of excess absorption ($N_{\rm H,excess}$, calculated at redshift zero) to column density 
($N_{\rm H,intrinsic}$, calculated at the redshift of the burst) is expected to depend on redshift approximately as
($1+z$)$^{2.6}$ (Galama \& Wijers 2001, see also Fig. \ref{nhvsz}).
This naturally led to the suggestion that excess X-ray absorption could be used as
some kind of redshift indicator, at least in the sense that bursts with high excess would
be expected to be at low redshift \citep{grupe07}.
In practice, this indicator has proven of limited value, in part because in many cases the measurement uncertainties
are quite high, but more pertinently for this work, also because many higher redshift
bursts exhibit surprisingly high excess absorption.

In fact, intrinsic X-ray column density appears to correlate strongly with redshift for the {\it Swift} GRB sample, as noted in previous works, most recently Campana et al. (2010), Behar et al. (2011), Campana et al. (2012) and Watson \& Jakobsson (2012), and as we show in Fig. \ref{nhvsz}.
At face value, this would seem to suggest GRBs were forming in increasingly dense environments
at higher redshift. Indeed, if the average metallicity of star-forming galaxies
were lower at
high-redshift, then the gas columns inferred would be greater still, since, as mentioned
above, the absorbing effect of metals is dominant over that of hydrogen and helium
except for very low metallicities.

Behar et al. (2011, hereafter B11) proposed that
the X-ray flux from distant objects like GRB afterglows and quasars 
suffers photo-electric absorption from the intergalactic medium (IGM). They tested a cold (neutral) diffuse IGM model and 
found that at high-redshift this could produce the dominant excess X-ray absorption component, providing
the metallicity was on average quite high, corresponding to a value at redshift zero of $\sim0.2-0.4$\,Z$_{\odot}$. 
At lower redshifts
they concluded that significant absorption intrinsic to the host galaxy is also required. 

Campana et al. (2012) concluded that the contribution of line-of-sight intervening systems to GRBs could produce the observed $N_{\rm H} - z$ relation, whereas Watson \& Jakobsson (2012) found that interveners alone are likely to be insufficient to account for the large columns seen at high-redshift. Both works find a correlation between X-ray absorbing column and optical `darkness' (see also Fynbo et al. 2009), showing that dark bursts tend to have large X-ray absorbing columns. If these bursts were being preferentially missed at low redshifts, they argue that this could explain at least some of the observed correlation between $N_{\rm H}$ and $z$. 

In this paper we re-visit the X-ray absorption seen in 198 {\it Swift} GRBs with known redshifts, and
consider whether there are any selection effects or systematic errors which could give rise
to the apparent increase of $N_{\rm H,intrinsic}$ with redshift.  
We also examine a large sample of AGN and quasars to see whether they are
consistent with the GRB findings.
Our conclusion is
that the $N_{\rm H,intrinsic}$--$z$ correlation appears to be real and therefore demanding of a physical explanation. 
We then perform a more realistic treatment of possible IGM absorption considering a range of temperatures and metallicities, and discuss the role of the observed cold Lyman-$\alpha$ forest and discrete intervening absorbers.
Finally we consider the alternative possibility, that the increasing column density is
intrinsic to the host galaxies at high-redshift, possibly due to changes in the mode of star formation
at earlier cosmic times.
Understanding the origins of X-ray absorption in GRBs and its evolution with cosmic time has the potential to provide a step forward in our understanding of the star formation process and/or the the missing baryon problem.

We use a flat Universe cosmology of $H_0 = 71$ km s$^{-1}$ Mpc$^{-1}$, $\Omega_{\rm M} = 0.27$, $\Omega_{\Lambda} = 0.73$ throughout.

\section{The $N_{\rm H}$ -- redshift relation} \label{sec:nhvsz}
In this section we reproduce the $N_{\rm H} - z$ correlation using our own analysis of the most recent {\it Swift} data, paying
particular attention to the highest redshift GRBs.
We then address the question of whether the trend could be due simply to selection effects or systematic
problems with the method of analysis.

\begin{figure*}
\begin{center}
\includegraphics[width=10cm, angle=90]{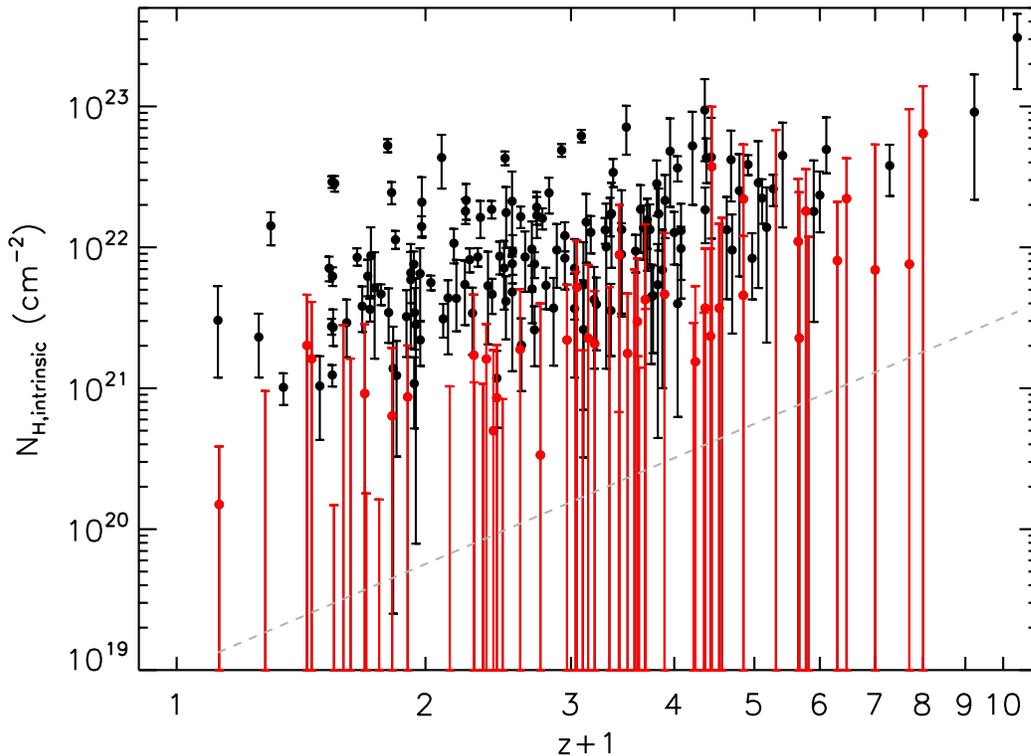}
\vspace{0.1cm}
\caption{Intrinsic X-ray column density, $N_{\rm H,intrinsic}$, as a function of redshift, $z$, as measured in absorbed power law fits to 198 {\it Swift} XRT-observed GRB afterglows up to 2012 September 1 with reported redshifts and PC mode spectra (see Footnote 1). Central values of the column density are represented by filled circles, and error bars are shown at the 90\% confidence level. Measured intrinsic absorption is denoted by black data points while upper limits are shown in red. An indication of the minimum detectable $N_{\rm H,intrinsic}$ with redshift is shown by the grey dashed line, for $N_{\rm H} = 10^{19}$ cm$^{-2}$.}
\label{nhvsz}
\end{center}
\end{figure*}

Our sample consists of all {\it Swift} XRT-observed GRB afterglows up to 2012 September 1
with reported redshifts (see Section \ref{redshifts}) and XRT Photon Counting (PC) mode X-ray spectra\footnote[1]{Taken from the UK {\it Swift} Science Data Centre repository at \url{www.swift.ac.uk/xrt_spectra} to ensure a homogeneous data reduction (Evans et al. 2009).}. This resulted in a sample of 198 sources whose X-ray spectra were modelled to obtain intrinsic absorber column densities.
We did not use data obtained in Windowed Timing mode to avoid, as far as possible, early times post-burst when spectral evolution is most likely to be occurring.
Data were processed with {\it Swift} tools 3.8 within HEASOFT 6.11. The latest calibration data at September 2012 (CALDB 3.9) were used and fitting was carried out in the X-ray spectral fitting package {\sc Xspec} version 12.7.0. We did not apply any binning to the spectra and have used Cash statistics (Cash 1979) as this is required for the spectra with the lowest count rates.

Each spectrum was modelled with a power law, absorbed by both a fixed Galactic component derived from the weighted average at the GRB position in the Leiden Argentine Bonn H\,I Survey (LAB Survey, Kalberla et al. 2005) and a component intrinsic to the source (redshifts are discussed in section~\ref{redshifts}). 
The models used for Galactic and intrinsic (or host galaxy) absorption were {\sc tbabs} and {\sc ztbabs} (Wilms et al. 2000) respectively with Solar metallicity, and the atomic data use abundances from Wilms, Allen \& McCray (2000) and cross-sections from Verner et al. (1996). We note that while Solar metallicity, Z$_{\odot}$, for the X-ray absorbing gas is conventionally adopted, X-ray absorbing column density scales approximately with the choice of metallicity.
We discarded seven bursts where the fits could not converge due either to poor signal to noise (GRBs 050509B, 061217, 071227, 100316D, 110918A and 120714B) or to the presence of an evolving blackbody component (GRB\,060218, e.g. Campana et al. 2006). 

The results are plotted in Fig. \ref{nhvsz}, and clearly show the same trend noted by previous authors.
The reduced sensitivity to absorption at higher redshifts (due to the fact that the softest photons are
shifted out of the observed energy band), is illustrated by the grey dashed line, however despite this, very few of the $z>4.5$ bursts only have an upper limit to their $N_{\rm H,intrinsic}$ measurements. The mean $N_{\rm H,intrinsic}$ for the 15 bursts at $z>4.5$, using the central, best-fitting value in all cases, 
is 4.5$\times$10$^{22}$ cm$^{-2}$.

The $N_{\rm H,intrinsic} - z$ correlation may indicate a lack of high-redshift GRBs for which no excess absorption above the Galactic value can be measured. This would then suggest that the progenitor stars of very high-redshift bursts had larger reservoirs of material around them either from the H\,{\sc II} region or a massive stellar wind. Alternatively it could be largely driven by absorption by gas in the IGM integrated over the entire path from the GRB to ourselves.
Before discussing these possibilities in more detail, we will investigate further the two highest redshift GRBs to examine the robustness of the very high intrinsic column densities observed in these cases. We then consider whether any biases or selection effects could be responsible for the apparent correlation.

\section{Further investigation of the highest redshift GRBs} \label{sec:seds}
GRBs 090423 and 090429B are the two highest redshift GRBs known, and both have high measured intrinsic X-ray column densities. We can investigate the robustness of their X-ray column measurements through the addition of near infrared (NIR) data. Combination of NIR photometry with X-ray spectra in a spectral energy distribution (SED), fitted in count space, provides a further, typically more accurate, method of X-ray column measurement (e.g. Starling et al. 2007). We can also infer a range of metallicities for these environments thereby testing the possibility of an origin in Population III collapsars. 

The SEDs for both GRB afterglows were fitted using the {\sc ISIS} spectral fitting package (Houck \& Denicola 2000) with power law and broken power law models. These intrinsic spectra are absorbed in the X-rays at our own galaxy and intrinsically to the host. The continuum is reddened in the NIR bands by Galactic extinction following the Milky Way extinction curve (Pei 1992) and in the host galaxy using a Small Magellanic Cloud (SMC) extinction curve (Pei 1992, often found most appropriate for GRB hosts e.g. Schady et al. 2007, 2012).
All absorption and extinction we measure here is along a single line-of-sight to the GRB. Galactic X-ray absorption values are taken from the LAB Survey (Kalberla et al. 2005) adopting the weighted interpolation of the nearest measured columns, while the Galactic extinction adopted for the NIR bands is taken from Schlegel et al. (1998). The atomic data use abundances from Wilms, Allen \& McCray (2000) and cross-sections from Verner et al. (1996). The following subsections detail this process and present the results: the temporal evolution of the GRBs is shown in Fig. \ref{lightcurves}, the SEDs and best-fitting models for both GRBs are shown in Fig. \ref{sedbestfit} and listed in full in Table \ref{tab:SEDfits}.

\subsection{GRB\,090423} \label{sec:423}

\begin{figure}
\begin{center}
\includegraphics[width=6.2cm, angle=90]{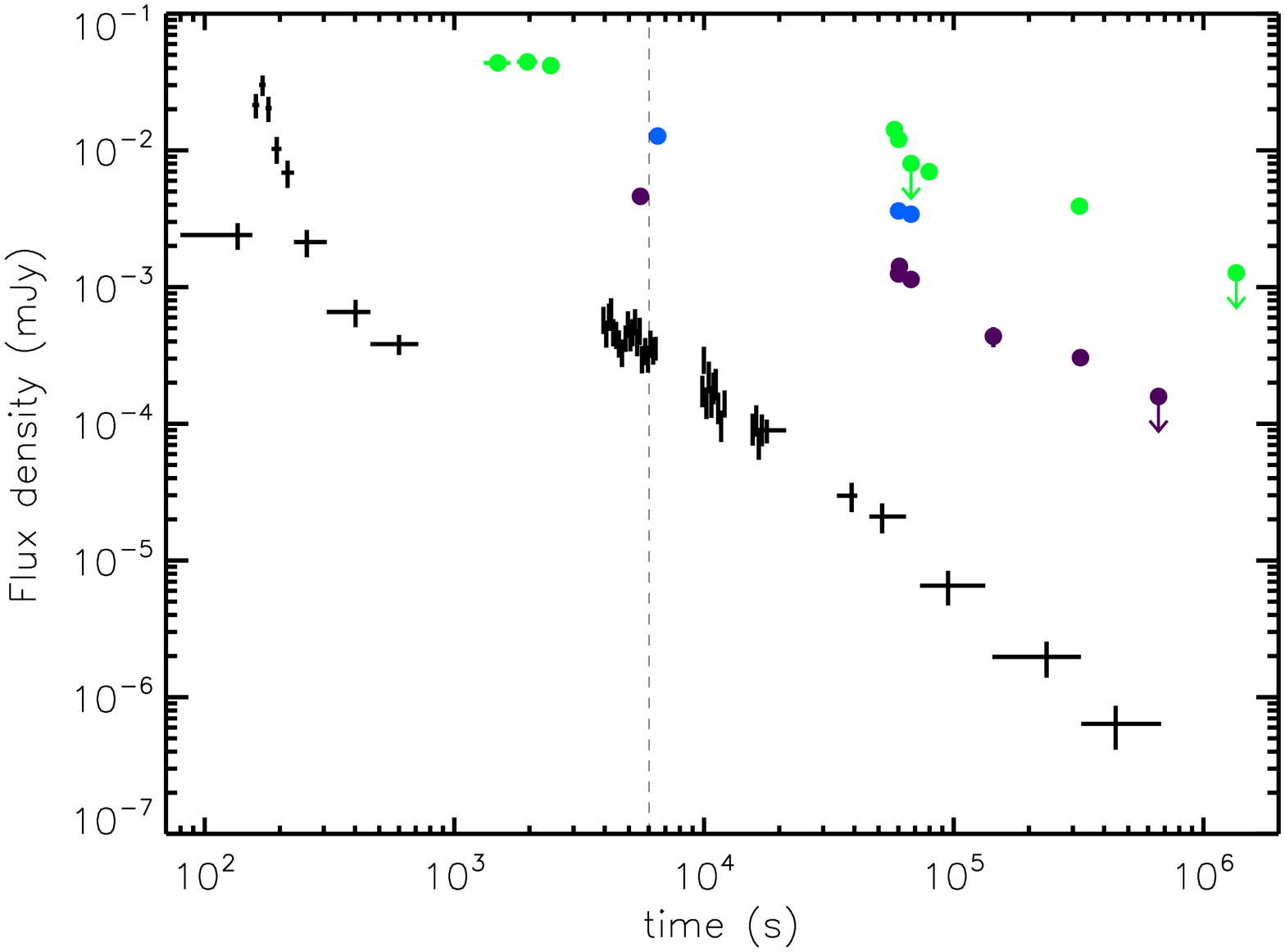}
\vspace{0.2cm}
\includegraphics[width=6.2cm, angle=90]{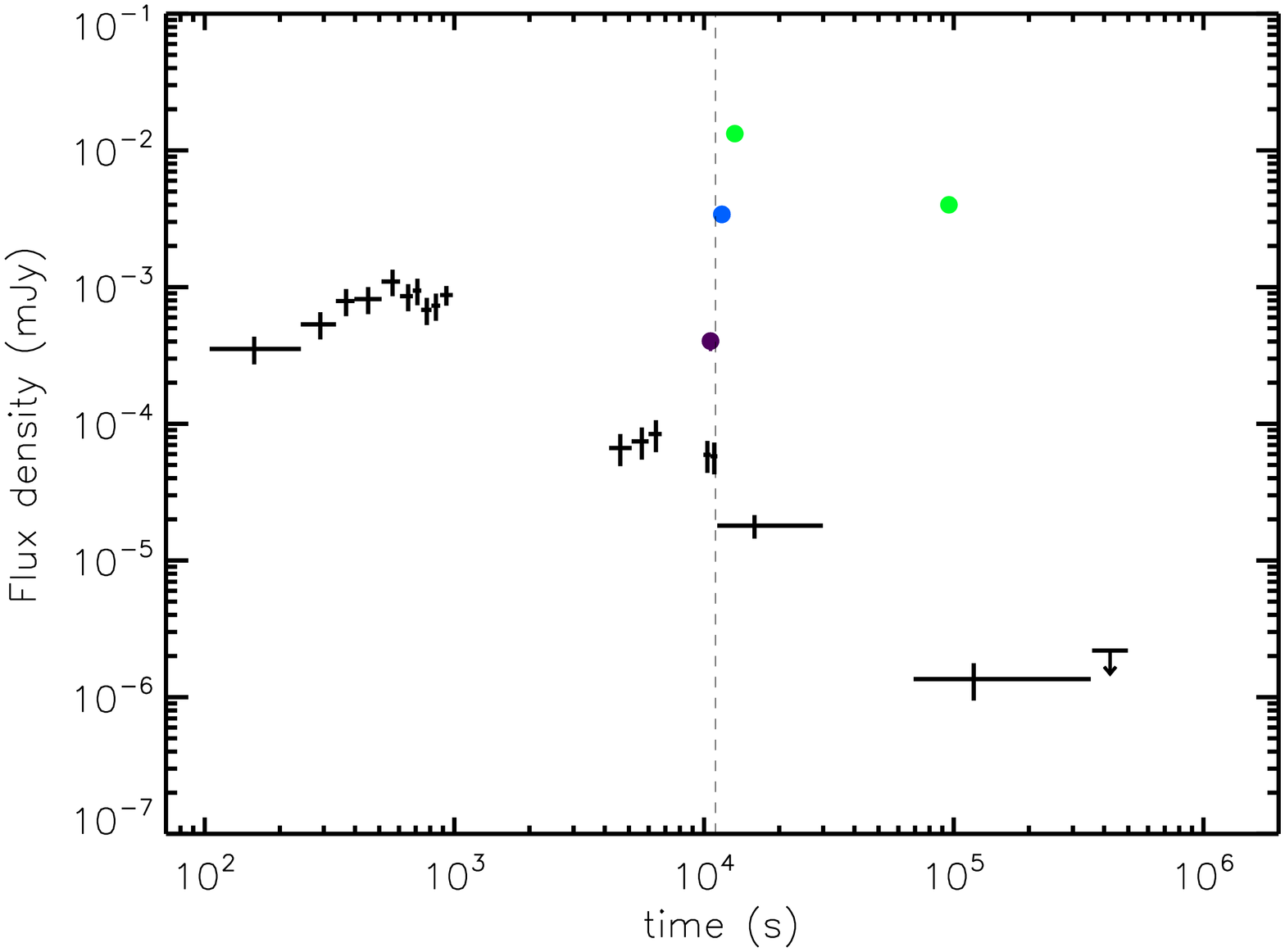}
\caption{Swift XRT X-ray (black) and $J$ (purple), $H$ (blue) and $K$ (green) observed light curves for GRB\,090423 (top panel) and GRB\,090429B (bottom panel). The times of SED creation are marked by vertical dashed lines. The specific X-ray flux at 2\,keV is shown, using the photon spectral index derived from X-ray-only fits. $J$ and $H$ band fluxes have been rescaled to $J$/7 and $H$/3 to clearly seperate them on the plot.}
\label{lightcurves}
\end{center}
\end{figure}

\noindent{\bf Observations}

GRB\,090423 was detected by the {\it Swift} Burst Alert Telescope (BAT, Barthelmy et al. 2005) on 2009 April 23. The narrow-field instruments, the X-ray Telescope (XRT, Burrows et al. 2005) and UltraViolet-Optical Telescope (UVOT, Roming et al. 2005), slewed and began observations 73\,s after the BAT trigger. The X-ray emission underwent an initial flare before settling to a constant spectral shape and rate of decay at $\sim$2000\,s after the BAT trigger, as shown in Fig. \ref{lightcurves}. The {\it Swift} observations are further described in Tanvir et al. (2009), Salvaterra et al. (2009) and Krimm et al. (2009).

NIR photometry was performed at Gemini North with the Near InfraRed Imager and Spectrometer (NIRI) instrument taking 9$\times$60\,s exposures in the $J$ and $H$ bands and calibrating to three 2MASS field stars.
We applied a correction to the $J$ band flux to account for the 2.8\% reduced transmission expected due to the intervening Lyman forest (Madau 1995; Curran et al. 2008).

We created an SED at time T$_0$+6020 seconds, chosen to coincide with the log mean time of the $J$ and $H$ band photometric observations at Gemini North (5560 and 6520 s respectively). There is a $K$ band data point close in time, but as the decay rate is not known we start by using $J$ and $H$ only, and then repeat our analysis including the $K$ band and assuming the extrapolation follows a constant decay as seen in our NIR light curve in Fig. \ref{lightcurves}. Following the method of Starling et al. (2007) we created a count space SED.

The X-ray data showed no spectral evolution from 2000\,s onwards, so to improve signal-to-noise we used a spectrum spanning a longer time-frame of 3910\,s to 6$\times$10$^5$\,s since trigger (exposure time 82000\,s),
well past the time of the X-ray flare at 300\,s, and normalised it to the count rate of a narrowly time-sliced spectrum with mean photon arrival time 6020\,s.

\begin{figure*}
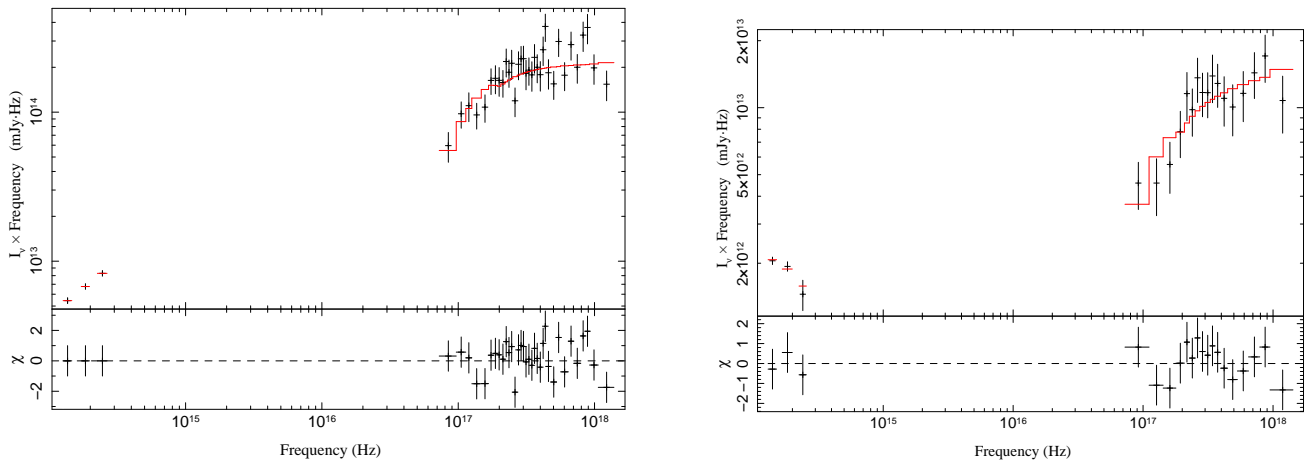

\begin{center}
\includegraphics[width=6cm, angle=-90]{figure3a.ps}
\hspace{1cm}
\includegraphics[width=6cm, angle=-90]{figure3b.ps}
\caption{SEDs for GRB\,090423 at 6020\,s (left, including the extrapolated $K$ band data point) and GRB\,090429B at 11112\,s and assuming $z = 9.4$ (right), plotted as $\nu F_{\rm \nu}$ vs. $\nu$ and on the same scales. The best fitting models are overlaid in red, consisting of an absorbed broken power law with slopes free for GRB\,090423 and a single power law for GRB\,090429B. In both cases intrinsic X-ray absorption was free to vary while intrinsic optical extinction was a variable parameter for GRB\,090429B but has been fixed to zero in fits to GRB\,090423 (see Sections 
\ref{sec:423} and \ref{sec:429}, and Table \ref{tab:SEDfits}).
The lower panels show the data to model ratio.}
\label{sedbestfit}
\end{center}
\end{figure*}

\noindent{\bf Results}

We find an absorbed power law is not an acceptable fit, with $\chi^2$/dof = 65.9/32. A broken power law is required ($\chi^2$/dof = 38.0/30), and we fit this both with the photon indices either side of the break free and with the difference in the indices fixed to 0.5 as expected for a synchrotron cooling break (e.g. Sari et al. 1998). Extinction in the host over the NIR bands was found to be negligible, tending to zero in all free fits; to reduce the number of free parameters with a view to obtaining the best possible constraints on $N_{\rm H,intrinsic}$, we fixed $E(B-V)_{\rm intrinsic}$ to zero in subsequent fits. This result also means that we cannot distinguish between optical extinction laws and is consistent with the findings of Zafar et al. (2011). Artificially introducing some optical extinction could not remove the X-ray column density we measure, but would in fact increase it further.  

Intrinsic X-ray absorption is detected at the 3$\sigma$ level, at $N_{\rm H,intrinsic}$ = (16$^{+11}_{-9}$)$\times$10$^{22}$ cm$^{-2}$ (Solar metallicity, broken power law with slopes free, optical extinction set to zero). This is consistent with the column density reported from X-ray analysis alone by Salvaterra et al. (2009) and the two values reported from the X-ray spectra presented in Tanvir et al. (2009). These previous works use differing atomic data than adopted here: for consistency we repeated our best-fit with these alternative atomic data \citep{Anders,Balucinska1,Balucinska2} and derived $N_{\rm H,intrinsic}$~=~($10^{+7}_{-6}$)$\times10^{22}$ cm$^{-2}$. 

We then went on to investigate the possibility of lower metallicity down to that of the LMC and then SMC (0.33 and 0.125\,Z$_\odot$ respectively, representative of most GRB hosts e.g. Schady et al. 2007,2012), 0.01\,Z$_\odot$ (e.g. Ledoux et al. 2009) and finally zero to represent Population III stars.
As expected, at lower metallicities the column density required to reproduce the observed absorption increases.  However, even at zero metallicity, the data can still be fitted thanks to the weak absorption of H/He, but only with a very high column. The implications of the zero metallicity fit are discussed further in Section \ref{sec:popiii}.

Finally, we extrapolated the 3 nearest $K$-band photometric data points listed in Tanvir et al. (2009) to 6020\,s and took the mean of these to add to the SED. Fitting the X-ray+$J$+$H$+$K$ SED with the model best-fitting the X-ray+$J$+$H$ SED we find similar results for the two, with no reduction in the uncertainties from adding in the $K$ band except in the upper limit on host galaxy extinction, $E(B-V)_{\rm intrinsic}$, when left to vary.

\subsection{GRB\,090429B}\label{sec:429}

\noindent{\bf Observations}

GRB\,090429B was detected by the {\it Swift} BAT on 2009 April 29. The XRT and UVOT slewed and began observations 106.6\,s after the BAT trigger. The X-ray light curve shows an initial rise, peaking between 500\,s and 600\,s after the BAT trigger (Fig. \ref{lightcurves}). Subsequent temporal coverage as the source faded is sparse. The {\it Swift} observations are reported in detail in Ukwatta et al. (2009).

NIR photometry was performed at Gemini North with the NIRI instrument in the $J$, $H$ and $K$ bands respectively (Cucchiara et al. 2011). We applied a Lyman forest correction to the $J$ band to account for the 64.9\% reduced transmission
of the emitted flux through the intervening Lyman forest (Madau 1995; Curran et al. 2008) at the redshift of $z = 9.4$ proposed by Cucchiara et al. (2011).

We created an SED at time T$_0$+11112 seconds, close to the log mean time of the $H$ band photometric observation at Gemini North. $J$, $H$ and nearest $K$ detections were transformed to T$_0$+11112\,s using the decay rate $\alpha = 0.53$ measured from the $K$ band in Cucchiara et al. (2011). This shift is very small and our results are not sensitive to this. We created an SED in an identical fashion to that created for GRB\,090423 (Section \ref{sec:423}). The X-ray data show no obvious spectral evolution so to improve signal-to-noise we used a normalised spectrum comprising all the X-ray data collected with {\it Swift} XRT, spanning the time-frame of 100\,s to 3$\times$10$^4$\,s since trigger (exposure time 8999\,s).

\noindent{\bf Results}

The same sequence of models applied to GRB\,090423 were fit to GRB\,090429B, adopting Solar metallicity for the intrinsic absorption. Single power law and broken power law models provide equally good fits to the data, and the F-test shows that the broken power law model does not provide a significant improvement. In the broken power law model the high energy power law slope and the break energy are unconstrained. Therefore we adopt the single power law model as the best representation of these data, with $\chi^2$/dof = 12/16. We find that, should it be located at $z=9.4$, there is a similar column of X-ray-absorbing gas in GRB\,090429B as determined for GRB\,090423, of $N_{\rm H,intrinsic}$ = (24$^{+16}_{-11}$)$\times 10^{22}$ cm$^{-2}$ (or (15$^{+9}_{-7}$)$\times 10^{22}$ cm$^{-2}$ using alternative atomic data for comparison with estimates presented elsewhere). 

Optical extinction is required in the single power law model fit at the level of $E(B-V)_{\rm intrinsic} = 0.02-0.07$, consistent with the results of Cucchiara et al. (2011) for the NIR data alone. Using, instead, a Milky Way extinction law we find that no distinction between extinction models can be recovered.

\subsection{Comparison with other $z>6$ GRBs}

We confirm the presence of a non-zero excess X-ray-absorbing column of gas in the host galaxy of GRB\,090423, consistent with that found in Salvaterra et al. (2009) and Tanvir et al. (2009) while using the different, typically more accurate method of count rate fitting to the broadband SED (e.g. Starling et al. 2007). Using the same technique for GRB\,090429B we find that, should the source be located at a redshift of $z=9.4$, a similar column of X-ray-absorbing gas is also required with a value consistent with that found by Cucchiara et al. (2011) for the X-ray data alone. 

In our sample we have four further $z>6$ GRBs among which only one has a constrained non-zero intrinsic column density. GRB\,050904 at $z = 6.29$ has, in our fits to its PC mode X-ray spectrum, a column density of (3.8$\pm$1.5)$\times$10$^{22}$ cm$^{-2}$. Significant work has been done on this source, including the creation of SEDs, largely due to the far higher quality of available data, hence we do not create an SED for this source here. Campana et al. (2007a) require metals to be present in its X-ray absorber from modelling of the observed evolution of the column density, which is initially of the same magnitude as that of GRB\,090423. They determine a metallicity limit of $Z \ge 0.03$\,Z$_{\odot}$, in-line with estimates from optical observations of GRB\,050904 (Kawai et al. 2006). GRB\,080913 lies at a redshift of $z = 6.7$, but for this afterglow the X-ray spectrum has too low signal to noise to be able to measure the X-ray column to adequate precision, even when including ground-based data (Greiner et al. 2009). In our fits to the PC mode X-ray spectrum the column density can only be constrained to lie below 9.6$\times$10$^{22}$ cm$^{-2}$. GRB\,100905A is tentatively included in this sample using the photometric redshift of $z = 7$ found with the United Kingdom Infrared Telescope (Im et al. in preparation, priv. comm.), resulting in a column density of (6.4$^{+7.5}_{-6.4}$)$\times$10$^{22}$ cm$^{-2}$. The final high-redshift source is GRB\,120521C at $z \sim 6$ (Tanvir et al. 2012a), for which we measure an intrinsic column density of $\le 5.4\times$10$^{22}$ cm$^{-2}$. The mean central value of all six $z>6$ GRBs in our sample is $N_{\rm H,intrinsic}=9 \times 10^{22}$~cm$^{-2}$. 

We note a further high-redshift candidate, GRB\,120923A. This source is suggested to lie at $z \sim 8$ due to a likely break between the Y and J-band fluxes (Levan et al. 2012). It is not included in our sample due to insufficient counts collected with {\it Swift} XRT to obtain reasonable constraints on any spectral fit. 

\section{Observational biases and selection effects}
\begin{figure*}
\begin{center}
\includegraphics[width=12cm, angle=90]{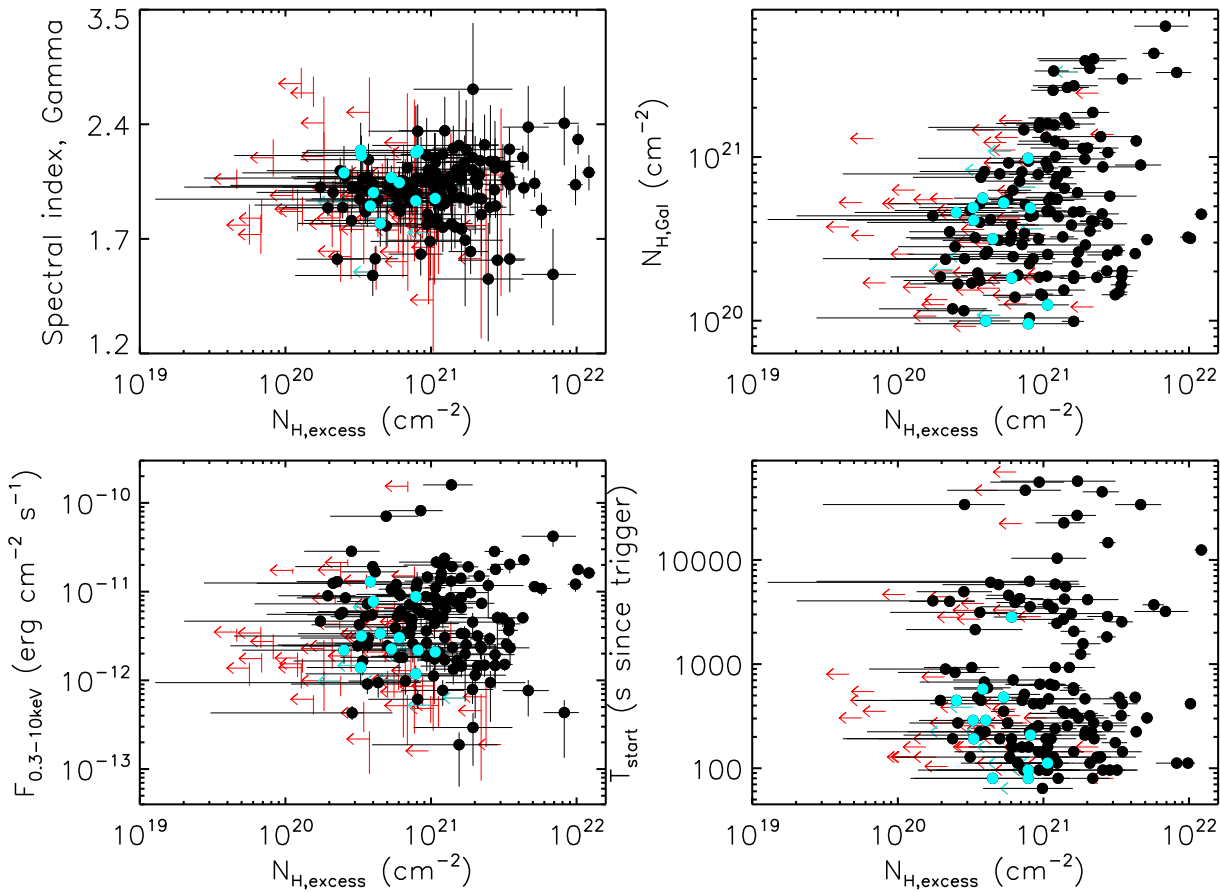}
\hspace{0.1cm}
\caption{Here we plot (across from top left to bottom right) the excess column density above the Galactic value at $z=0$, $N_{\rm H,excess}$, against spectral index, Galactic column density, integrated X-ray flux and observation start time to diagnose any biases in the data or fitting method. Measured column densities are shown in black and upper limits are shown in red, while we highlight $z \ge 4$ GRBs in cyan.
}
\label{checks}
\end{center}
\end{figure*}

Having confirmed the presence of high X-ray absorbing columns in our highest redshift GRBs, we now examine factors which may affect the observed column densities of the sample as a whole.
In Fig. \ref{checks} we show the relationship between excess X-ray absorbing column, $N_{\rm H,excess}$, and various characteristics of the GRB X-ray spectra. We define $N_{\rm H,excess}$ as the column density required over and above that of the Galactic column, when redshift is set to zero (as fitted in Section \ref{sec:nhvsz} using $z=0$). From this we may ascertain the level of any degeneracies between fitted parameters, and effects of spectral evolution on our results without any redshift dependencies. The $z \ge 4$ GRBs are highlighted in cyan in Fig. \ref{checks}. 
In the following subsections we examine the most relevant of these and other potential contaminants to an intrinsic $N_{\rm H} - z$ relation.

\subsection{Redshifts} \label{redshifts}

In all our fits we have adopted the redshifts of the GRBs primarily from Fynbo et al. (2009), Jakobsson et al. (2012), Kr\"uhler et al. (2012) or the {\it Swift} XRT GRB repository$^1$ and references therein. 
The majority (97\%) of these redshifts are spectroscopically determined and hence very accurate, and photometric redshifts are used only for GRBs 100905A (Im, priv. comm.), 090429B (Cucchiara et al. 2011) and 050814 ($z = 0.53\pm 0.3$, Jakobsson et al. 2006) and the GROND$+${\it Swift} UVOT photometric redshift sample (Kr\"uhler et al. 2011) comprising GRBs 080825B ($z=4.31^{+0.14}_{-0.15}$), 081228 ($z=3.44^{+0.15}_{-0.32}$), 081230 ($z=2.03^{+0.16}_{-0.14}$) and 090530 ($z=1.28^{+0.15}_{-0.17}$). If we remove the sources with photometric redshifts from the sample we still find a correlation between $N_{\rm H,intrinsic}$ and $z$. For the high-redshift source GRB\,090423 there is an error associated with the redshift of $z = 8.23^{+0.07}_{-0.06}$ (see Tanvir et al. 2009 for details of how this was determined). Using the upper and lower limits on $z$ in our SED fits, no appreciable difference was found in the results, e.g. the X-ray column density changes by no more than 3\%. The redshift of GRB\,090429B is less secure, no spectroscopic information being available, and the $z=9.4$ redshift determination is detailed in Cucchiara et al. (2011).

\subsection{Galactic column density} 

In our earlier analysis we did not consider the errors associated with the assumed Galactic foreground column density. The errors on these column densities are not explicitly given in the LAB Survey, but are likely to be of order 10\%, significantly smaller than the excess column we find in most GRBs. Fig. \ref{checks} shows the Galactic versus measured excess column density for each of our sample GRBs when fitting at zero redshift. The dearth of data points in the top left of the $N_{\rm H,Gal} - N_{\rm H,excess}$ distribution illustrates that as Galactic column increases, the lower limit on any measurable excess column increases with it. However, the fraction of upper limits on excess column among all data points does not increase and while this may be due to low number statistics (with only 19\% of the sample intercepting a Galactic column density of 10$^{21}$ cm$^{-2}$ or above), it is suggestive of a bias in the distribution. 

\begin{figure*}
\begin{center}
\includegraphics[width=10cm, angle=90]{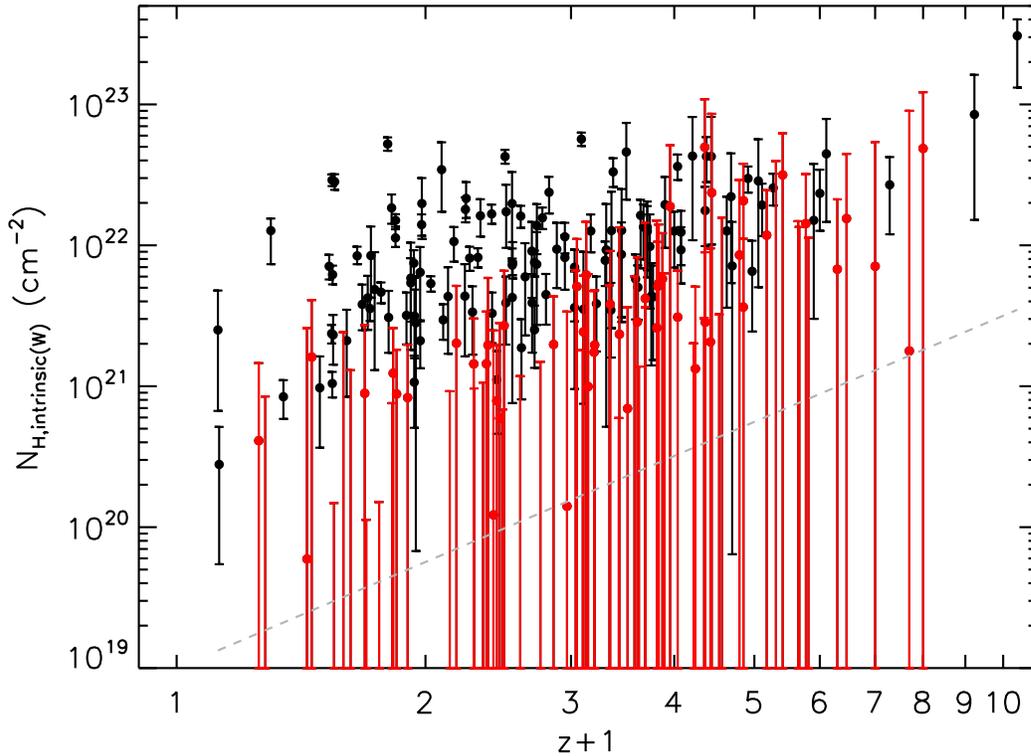}
\vspace{0.1cm}
\caption{The relation between $N_{\rm H,intrinsic(W)}$ and redshift, $z$, when Galactic column includes a fractional contribution from H$_{\rm 2}$ as defined in Willingale et al. (2013). This figure can be directly compared with Fig. \ref{nhvsz}, and uses the same symbols, colour coding and axis ranges.
}
\label{withMWH2}
\end{center}
\end{figure*}

\begin{figure}
\begin{center}
\includegraphics[width=5cm, angle=90]{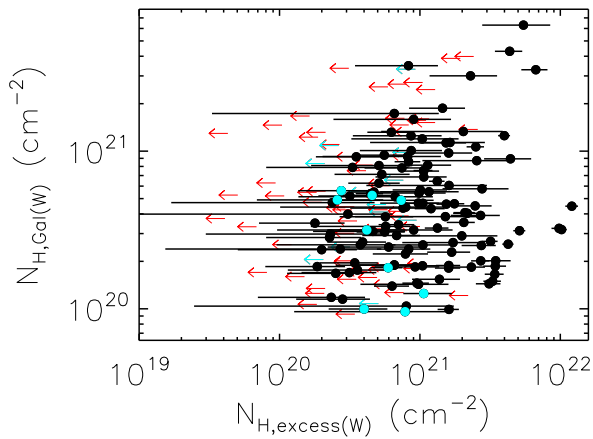}
\vspace{1cm}
\caption{Excess column density above the Galactic value at $z=0$, $N_{\rm H,excess(W)}$, against Galactic column density where Galactic column includes a fractional contribution from H$_{\rm 2}$ as defined in Willingale et al. (2013). This figure can be directly compared with the top right panel of Fig. \ref{checks}, and uses the same symbols, colour coding and axis ranges.
}
\label{withMWH2_checks}
\end{center}
\end{figure}

If H\,I column is not accurately translated into an X-ray equivalent hydrogen column in regions of high density, an underestimate, rather than an overestimate, of the true value is expected if H\,I represents some but not all of the material present. Willingale et al. (2013) point out that the molecular hydrogen contribution is assumed to be 20\% of the input value in the X-ray absorption model we use ({\sc tbabs}, Wilms et al. 2000) and as such, use of the LAB H\,I values is incorrect. Instead, they derive an empirical direction-dependent measure for Galactic H$_{\rm 2}$ fraction using both the atomic hydrogen column density and the dust extinction in the Galaxy. Following their method, we can refit our GRB sample with the revised, fixed Galactic column densities to derive intrinsic absorption. The resulting $N_{\rm H(W)} - z$ relation and correlation of various parameters are shown in Fig. \ref{withMWH2} (to compare with Fig. \ref{nhvsz}). We distinguish these column densities from previous measurements with the suffix (W). While some of the measured intrinsic columns are now visibly different, the majority show no appreciable change and the $N_{\rm H} - z$ relation remains. There is a marked change in the $N_{\rm H,Gal} - N_{\rm H,excess(W)}$ distribution, shown in Fig. \ref{withMWH2_checks} (to compare with the top right panel of Fig. \ref{checks}), such that no significant correlation is now present (as expected and shown for a subset of these data in Willingale et al. 2013).  

We distinguish the highest redshift GRBs ($z \ge 4$) among our sample in cyan in Fig. \ref{checks}, and note that only two of these (at $z = 4.31,4.65$) suffer from especially high Galactic absorption and in both no excess column was measured. 
In Sections \ref{sec:423} and \ref{sec:429} we discussed choices for the value of the Galactic column density and the effect this has upon the SED fits of the two highest redshift GRBs. We note that there is no effect on the resulting intrinsic column when switching between weighted mean or nearest measured Galactic column density in SED fits to GRBs 090423 and 090429B. The inclusion of H$_{\rm 2}$ in the Galactic column values as above results in only a small increase in these cases of $N_{\rm HI,LAB} \times 1.086$ for 090423 and $N_{\rm HI,LAB} \times 1.040$ for 090429B, on their already low Galactic H\,I columns of 3 and $1\times$10$^{20}$ cm$^{-2}$ respectively.
Thus we can conclude that uncertainties in the Galactic columns, and specifically an incorrect contribution from H$_{\rm 2}$, will not remove the correlation we see between intrinsic X-ray column and redshift.


\subsection{Atomic data} 

Several atomic data tables of (Solar) abundances and photoionisation cross-sections are available for use with X-ray spectral models. We have used abundances from \cite{Wilms} and cross-sections from \cite{Verner}.
Our X-ray column density measurements are consistently higher than those reported in the {\it Swift} XRT GRB spectral repository$^1$ due to the use of differing atomic data. We note that this results in some cases to detections via our method where only an upper limit was derived in the XRT repository fits. In particular this affects sources for which the Galactic column density is very high, since the same H\,I value translates to a lower X-ray column density in our method, leading to a higher required excess X-ray column density.   
In all our fits we have used what we believe to be the most up to date atomic data available within the {\sc Xspec} spectral fitting package and those that are recommended for use together with the {\sc tbabs} absorption model. To enable comparison with previous works on GRBs 090423 and 090429B we repeated our SED fits using alternative atomic data and Table \ref{tab:SEDfits} shows that this results in a decrease in the X-ray column densities by a factor $\sim$1.65 (see also Watson 2011), while remaining consistent within the 90\% errors and recovering an excess above the Galactic value in both cases. 

\subsection{Spectral evolution} 

The underlying continuum must be well understood in order to model the absorbing columns. GRB afterglows emit synchrotron radiation which manifests as a simple power law in the X-ray band with typical photon index between 1.5 and 2.5. Spectral evolution is common in the early evolution of GRBs (e.g. Zhang, Liang \& Zhang 2007), hence we restricted our spectra to PC mode data which are taken when the source count rates fall to below about 1 count s$^{-1}$ when we expect most GRBs to have reached the afterglow phase. To get an indication of the affect of spectral evolution on our measurements we plotted spectrum start time against excess X-ray absorbing column in Fig. \ref{checks}, and, while the spacecraft constraints (orbital gaps and observing restrictions) are apparent, no correlation was found. Additionally, the measured photon indices all lie in the expected and previously observed (e.g. Evans et al. 2009) range so we can be confident that the underlying spectrum is reasonably accounted for.

\subsection{Metallicity evolution} 

There are some indications that the average metallicities of GRB hosts are a weak function of redshift (Fynbo et
al. 2006; Savaglio, Glazebrook \& Le Borgne 2009), in which higher redshift hosts are more metal poor. 
This would have the effect of increasing the equivalent hydrogen column densities at high redshifts for the X-ray absorption within the host galaxy, 
strengthening the $N_{\rm H,intrinsic} - z$ correlation. Unfortunately the relation between redshift and metallicity of GRB hosts is not well constrained,
mainly because of the relatively small number of reliable metallicity measurements (see e.g. Savaglio, Glazebrook \& Le Borgne 2009).
In addition, it is particularly difficult to measure the metallicity in the direct vicinity of the GRB, where the X-ray absorbers within the 
GRB host may be expected to lie: the gas producing the absorption lines in optical spectra (from which abundances are measured) is 
generally located hundreds of parsecs away from the burst site (as measured directly, see e.g. Vreeswijk et al. 2007, or indirectly, see e.g.
Schady et al. 2011). For these reasons, and because of the complex observing biases associated with optical spectroscopy of GRB afterglows (e.g. Jakobsson et al. 2012) we choose not to include a redshift dependent metallicity estimate in fits to the data, but we will assess its impact on the the absorption from intervening material for three evolutionary scenarios in Section \ref{sec:igm} and Fig. \ref{fig12}.

\subsection{A selection effect against low-redshift high-column density systems, or high-redshift low-column density systems?}

Campana et al. (2010) proposed that the the dearth of heavily absorbed GRBs at 
low redshift ($z \le 1-2$)
could be due to the difficulty in obtaining redshifts for these systems if the high gas column was
also accompanied by large dust extinction.
They argued that if such systems exist and were included, much of the evidence for a correlation between
 $N_{\rm H,intrinsic}$ and redshift would go away. With larger samples including a number of dark bursts, Campana et al. (2012) and Watson \& Jakobsson (2012) did find some low-redshift high-column density systems, but not enough to account for the entire correlation.
It is certainly plausible that such systems would be selected against.
Harder to understand is why they would appear in our sample at higher redshifts,
since there the effects of dust extinction should be amplified where
we observe rest-frame UV light in the optical. For all these reasons we do not feel that there are many low-redshift high-column density systems going undetected and which would remove the observed correlation.
Arguably GRBs with $z>2$ are relatively easy to
obtain redshifts for from their afterglows since they usually exhibit strong Ly$\alpha$ absorption, but in fact
in few of these systems is significant dust extinction seen (even in cases of high X-ray absorption, e.g. Zafar et al. 2011).
It would therefore seem to require a significant evolution in dust properties with redshift,
in the sense that the effective dust to metal ratio in GRB sight lines even at $z\sim2$, was
considerably lower than at $z\sim0$ as concluded by Watson \& Jakobsson (2012).

Alternatively, we might ask whether at high redshift we could be preferentially losing
low-column systems from the sample with redshifts. An argument could be that 
bursts in low density environments would show low $N_{\rm H,intrinsic}$ and
have relatively weak absorption lines (especially Lyman-$\alpha$ which is usually
very prominent), and hence appear as featureless continua in low S/N spectra.
While this could plausibly arise occasionally, it is very unlikely to be a major
effect, since the break at Lyman-$\alpha$ by $z\sim4$ is strong simply due to the
Lyman-$\alpha$ forest, even in the absence of strong host absorption.

\begin{figure*}
\begin{center}
\includegraphics[width=10cm, angle=90]{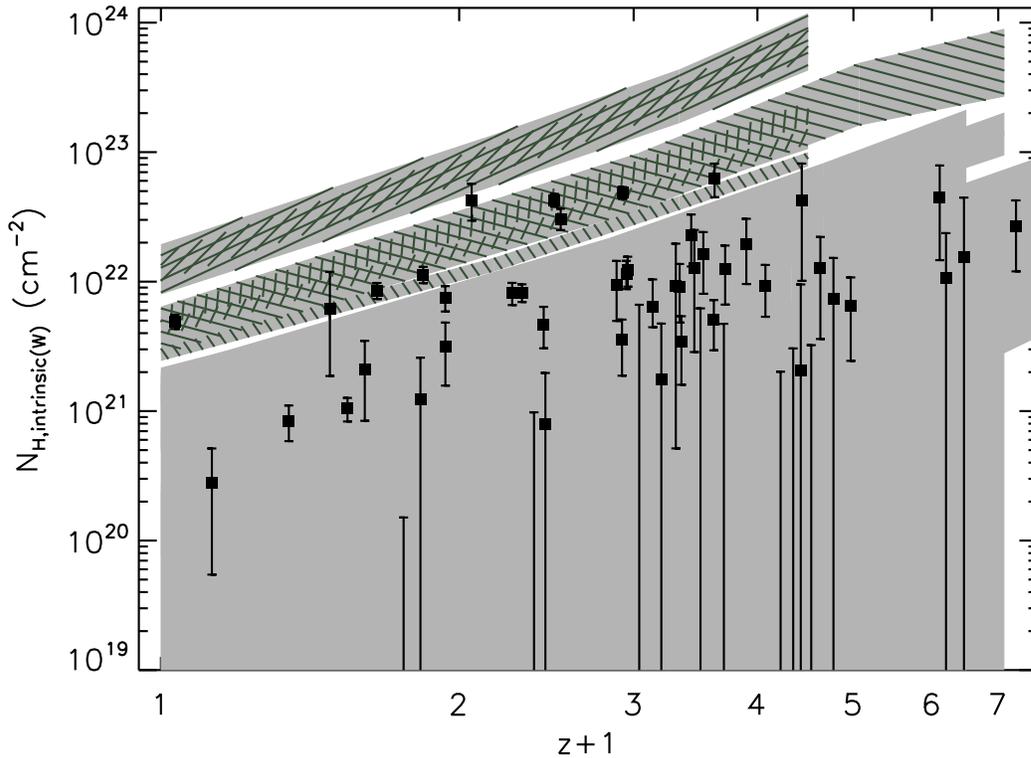}
\vspace{0.1cm}
\caption{Intrinsic X-ray column density as a function of redshift, as measured in absorbed power law fits to the TOUGH sample of GRBs (Hjorth et al. 2012; Jakobsson et al. 2012; Kr\"uhler et al. 2012). Central values are represented by filled squares, with 90\% confidence error bars shown. For the 16 sources with only an estimated redshift range we calculated $N_{\rm H,intrinsic(W)}$ for several redshifts in the allowed range and show these as grey regions containing the 90\% confidence intervals. Hatching indicates six individual bursts which could have large intrinsic column densities at low redshifts.}
\label{tough}
\end{center}
\end{figure*}

In order to investigate these questions further we analysed separately the TOUGH sample of Hjorth et al. (2012)
which consists of a subset of {\em Swift} GRBs, selected independently of their optical
properties, for which it was attempted to identify and obtain redshifts for the host galaxies, even for those
without a redshift or location from an optical afterglow. Thus this sample is optically unbiased, and
unusually complete, although 16 of the 69 bursts remain without redshifts and
only have constraints based on detection of afterglow or host in some optical band (and in one case there is not even that constraint).
We then fitted all GRBs in the TOUGH sample, using the time-averaged PC spectra$^1$ as we did for our redshift sample in Section \ref{sec:nhvsz} (given the completeness of the TOUGH sample, we included GRB\,060218, which lies at $z=0.0331$, by adopting the late-time PC spectrum).
Where only a redshift range was available we fitted at several values within the interval to produce the region in the $N_{\rm H,intrinsic} - z$ plane in which
the GRB is likely to lie (Fig. \ref{tough}). We used both LAB Survey Galactic column densities (as used in Section \ref{sec:nhvsz}) and H$_{\rm 2}$-corrected Galactic columns as described in Section~4.2 and no appreciable difference was found in the results for the sample as a whole: from hereon we have used the H$_{\rm 2}$-corrected column densities.
Visually it does not appear that the dearth of high-column density low-redshift
bursts is explicable by a selection effect, since if it were there 
should be more bursts without redshifts whose region of allowed $N_{\rm H,intrinsic}-z$ 
appeared in the upper left part of Fig. \ref{tough}.
We consider splitting the sample at $N_{\rm H,intrinsic}=10^{22}$~cm$^{-2}$, the approximate mean column density of the known-$z$ sample, and taking
the ratio of bursts above (high) and below (low) that column in two redshift ranges $z<1$ and $z>5$.
If we include all those bursts whose loci cross the region $N_{\rm H,intrinsic}>10^{22}$ and $z<1$ within the high column bin only,
then the {\it maximum} ratio of high to low columns is 9:19 in this low redshift range.
At high redshift ($z>5$) we find the {\it minimum} ratio of high to low columns is 6:5,
again where we have taken bursts of unknown redshift which at least might be $z>5$ and $N_{\rm H,intrinsic}<10^{22}$~cm$^{-2}$ range
and counted them in the low column bin only.
This supports the finding from the full sample that beyond $z=5$ the majority of GRB sight-lines have $N_{\rm H,intrinsic}>10^{22}$~cm$^{-2}$.

\begin{figure}
\begin{center}
\includegraphics[width=6.4cm, angle=90]{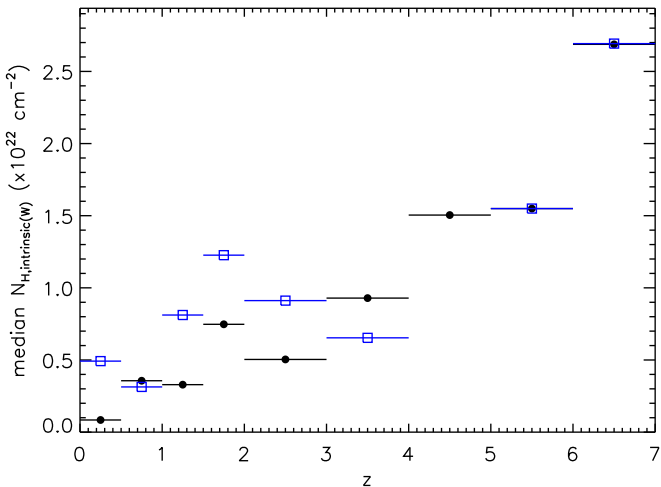}
\hspace{0.1cm}
\caption{The binned distribution of $N_{\rm H,intrinsic(W)}$ versus $z$ for the two GRB samples is shown here: {\it Swift} GRBs (black filled circles) and TOUGH GRBs with known redshifts (blue open squares).}
\label{histogram}
\end{center}
\end{figure}

In a comparison of the results for the two GRB samples (the full {\it Swift} redshift sample and TOUGH sample with known redshift) 
we show how the median $N_{\rm H,intrinsic}$ per redshift bin evolves over redshifts of $\sim$0--7 (Fig. \ref{histogram}). From the larger {\it Swift} GRB sample it can be seen that the $N_{\rm H,intrinsic} - z$ evolution occurs over all redshifts and not only in the highest redshift bins. The TOUGH sample consists of a greater percentage of low redshift sources, and again we see the median column at low redshift is substantially lower than at high redshift, while the two samples show similar though not identical evolution, despite a dearth of TOUGH Sample objects known to lie at $4<z<5$.

\section{A study of AGN and quasar samples} \label{sec:nhvszagn}
\begin{figure*}
\begin{center}
\includegraphics[width=10cm, angle=90]{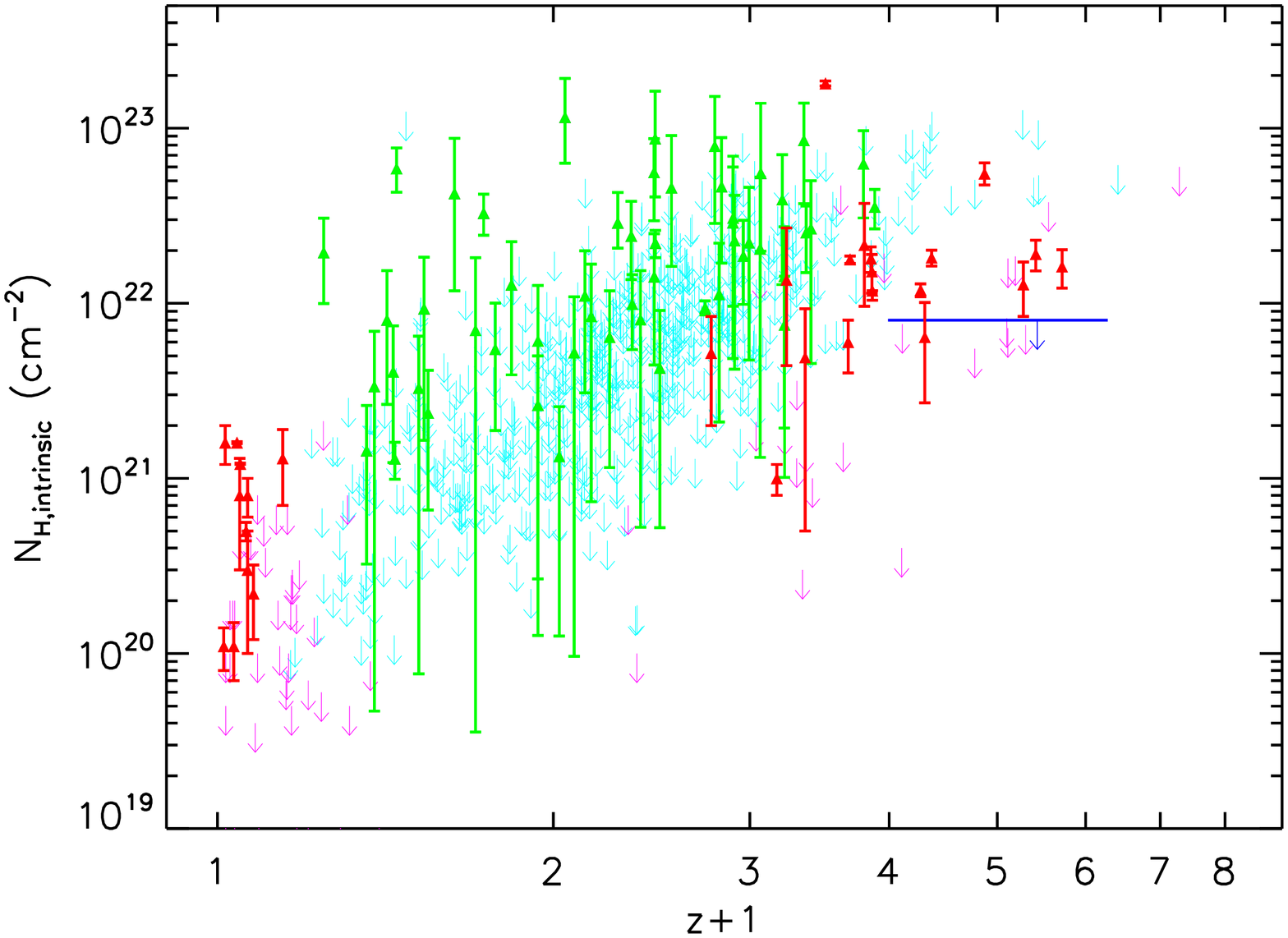}
\vspace{0.1cm}
\caption{Intrinsic X-ray column density, $N_{\rm H,intrinsic}$, as a function of redshift, $z$, as measured in absorbed power law fits to AGN and quasars. A sample of 678 AGN from Scott et al. (2011) are shown as green triangles and cyan upper limits. We overplot literature measurements (red) and upper limits (magenta) for quasars from both Page et al. (2005) and B11.
The upper limit from stacking spectra of 48 radio-quiet high-redshift quasars (Vignali et al. 2005) is shown in blue. All error bars and upper limits are shown at the 90\% confidence level.}
\label{nhvszagn}
\end{center}
\end{figure*}

We also applied the same absorbed power law model used for the GRBs to a subset of AGN taken from
the sample described in Scott et al. (2011). These sources are
selected from a cross-correlation of the SDSS DR5 quasar catalogue
(Schneider et al. 2007) and the second incremental \textit{XMM-Newton}
serendipitous X-ray source catalogue (2XMMi-DR2; Watson et al. 2009).
Objects which were not well fit (null hypothesis probability,
$p<1\%$) by any of the 4 simple models used in the analysis and
objects found to require a soft excess component in their best fit were
removed, leaving a subsample of 678 sources which are re-analysed
here. This new analysis finds 52/678 sources with a constrained non-zero
$N_{\rm H,intrinsic}$ component, plotted in Fig. \ref{nhvszagn} as green
triangles. The remaining sources are plotted as cyan upper limits.
We have also gathered measurements of $N_{\rm H,intrinsic}$ from other
samples in the literature in order to increase the redshift range
covered (hereafter referred to as quasars), although we note that the modelling may not have been carried
out in the same way as is done here for our GRB and AGN samples. Included in Fig. \ref{nhvszagn} are 94 quasars from Page et al. (2005) and B11, plotted as red triangles (for constrained $N_{\rm H,intrinsic}$ measurements) or magenta upper limits. Also shown (in blue) is the result from a mean spectrum of 48 $z>4$ quasars from Vignali et al. (2005). In fits to the AGN for Fig. \ref{nhvszagn} we adopted LAB Galactic column densities for comparison with the reported quasar fits.

Both the GRBs and AGN+quasars show a correlation of column density with redshift. A constant value does not provide an acceptable fit to either sample for sources with detectable column densities (weighting each point by its average 1$\sigma$ error).
Of course, some correlation in this plot is expected artificially, due to the fact that the detection threshold for
the minimum measurable column density increases approximately as $(1+z)^{2.6}$ (as shown in Fig. \ref{nhvsz}).
The question then is whether the combination of measured column densities and upper limits could be consistent 
with a non-evolving distribution, with an important contribution to the observed column densities coming perhaps from intervening absorbing material common to all extragalactic sources.

\section{Absorption from intervening material} \label{sec:igm}

The measured X-ray column densities of both AGN and GRBs may also comprise some contribution from the IGM, as recently pointed out by B11.
These authors computed this contribution for the case of a cold, neutral IGM, and we re-examine this and model the effects of a
more realistic warm-hot (ionised) diffuse IGM component. We then consider absorption by the known cold component along the line-of-sight, namely the 
Lyman-$\alpha$ absorption systems also seen in the optical spectra.

\subsection{Possible contribution of the WHIM}

The majority of baryons in the universe are thought to reside in the diffuse intergalactic medium,
in a range of warm to hot phases.
In particular, a large proportion of these IGM baryons are expected to exist in a warm-hot phase 
($\sim10^5-10^7$\,K), the so-called Warm-Hot Intergalactic Medium (WHIM). Although observational constraints
on its extent and properties are still very scarce, theoretical modelling suggests it is likely
to be moderately enriched in metals that have been expelled from galaxies \citep{Wiersma,Cen}.

The photo-electric optical depth (opacity) of the IGM of metallicity
relative to Solar,
$Z$, at observed photon energy $E$ for a source at 
redshift $z$ is given by Equation 3 of B11,

\begin{equation}
\tau_{\rm IGM}(E,z,Z)=\int_{0}^{z} n_{\rm H}(z')\sigma(E,z',Z) c \left( \frac{dt'}{dz'}\right) dz'.
\label{eq1}
\end{equation}

They consider a cold IGM and assume the cross-section per hydrogen atom 
scales with metallicity, 
$\sigma(E,z,Z)\propto Z\sigma(E,z)$.
If the IGM is cold the cross-section should be expressed as the sum of
two components, the hydrogen (H) and helium (He) which are not part of the
metallicity complement, and the higher atomic number elements which are.
We therefore rewrite Equation \ref{eq1} above as

\begin{equation}
 \tau_{\rm IGM}(E,z,Z) = \frac{n_{0}c}{H_{0}} \int_{0}^{z} (\sigma_{\rm HHe}(E,z')+\sigma_{\rm metals}(E,z',Z))
\label{eq2}
\end{equation}

\[
 ~~~~\quad\quad\quad\quad\quad\quad \cdot~\frac{(1+z')^{2}}{\sqrt{(1+z')^{3}\Omega_{\rm M}+\Omega_{\Lambda}}} dz'
\]

where $n_{0}$ cm$^{-3}$ is the IGM number density at $z=0$,
$H_{0}$ is the Hubble
constant, $\Omega_{\rm M}$ and $\Omega_{\Lambda}$ are, respectively, the
present-day matter and dark energy fractions of the critical energy
density of the Universe and $c$ is the velocity of light.
When the metallicity drops to zero, $Z=0$, the cross-section
of the metals disappears, $\sigma_{\rm metals}(E,z,Z)=0$, but
there is still opacity due to the hydrogen and helium cross-section,
$\sigma_{\rm HHe}(E,z)$.
B11 further assume that the cross-section
scales as $\sigma(E,z)\propto (1+z)^{-2.5}$.
As they point out, this is a reasonable approximation for cold material
but not for warm material.

\begin{figure}
\begin{center}
\includegraphics[height=8.2cm,angle=-90]{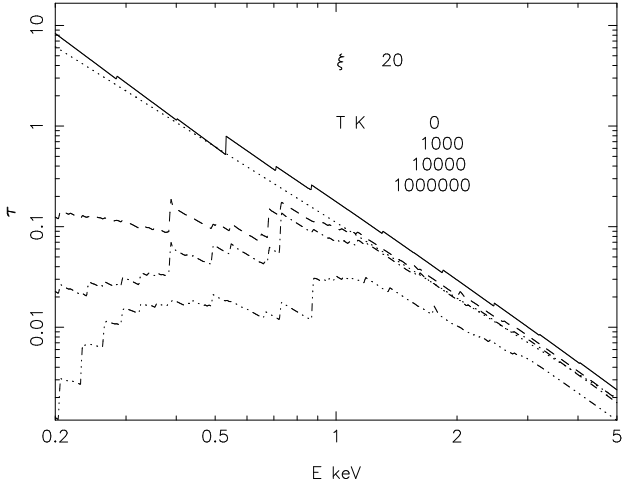}
\end{center}
\caption{The optical depth of the IGM at 0.5\,keV as a function of energy, for a column of $10^{21}$ cm$^{-2}$
with Solar abundances. The solid curve is a cold column using the model {\sc tbabs} in {\sc XSPEC}. The dotted curve is the approximation adopted by B11.
The dashed, dot-dashed and triple-dot-dashed curves are for warm columns using
the parameters given in the {\sc XSPEC} model {\sc absori}.}
\label{fig10}
\end{figure}

It is well established that the diffuse IGM is largely ionised (e.g. Gunn \& Peterson 1965; Dav\'{e} et al. 2001), and so for our purposes should be treated as a warm/hot absorber. To investigate the effects of such a component, we calculate the effective absorption for a range of simple, single phase, warm IGM models.
Fig. \ref{fig10} shows a comparison of the photo-electric absorption of
a cold ISM with Solar abundance calculated using the {\sc XSPEC} model {\sc tbabs}, the B11 approximation above and the absorption
of a warm absorber at 3 different temperatures estimated using the {\sc XSPEC} model {\sc absori} which is based on Done et al. (1992). We used the {\sc absori} model because it gives direct access to the photoionising spectrum. It includes absorption edges but not line opacity and therefore provides only a crude approximation to the ionisation balance. In the current context where we integrate over a large range of redshift this is unimportant because high resolution details in the absorption profile
are lost. The model is adequate to provide the sensitivity to the level of photoionisation and temperature.

The power law approximation used by B11 is normalised to the value at 0.5\,keV which
is just below the oxygen K absorption edge; this is why it is significantly
lower than the solid curve for all energies above this edge in Fig. \ref{fig10}. 
The ionisation parameter $\xi$ was estimated using the measured cosmic
X-ray background spectrum (CXRB, De Luca \& Molendi 2004; Fabian \& Barcons 1992). This is a power law with photon index
1.4 and normalisation 11.6 photons cm$^{-2}$ s$^{-1}$ sr$^{-1}$
keV$^{-1}$ in the energy range 1--10\,keV and falls off as a bremsstrahlung
spectrum at higher energies. The integrated CXRB flux is $F_{CXRB}\sim2.9\times10^{-7}$ ergs\,cm$^{-2}$ s$^{-1}$ sr$^{-1}$ over the energy band 0.005-300\,keV.
The ionisation parameter appropriate for the IGM is therefore
$\xi\approx4\pi F_{CXRB}/n_{e}$ where $n_{e}$ is the electron density. The electron density is a factor $\sim$1.2 higher than
the hydrogen density $n_{0}$ which we take from
B11 (and references therein) to be $n_{0} \approx1.7\times10^{-7}$ cm$^{-3}$. This yields $\xi\approx20$\,ergs\,cm\,s$^{-1}$,
which is the value assumed in Fig. \ref{fig10}. For this ionisation parameter and temperatures of $10^{3}-10^{6}$\,K the absorption at low energies is significantly reduced. 

\begin{figure}
\begin{center}
\includegraphics[height=8.4cm,angle=-90]{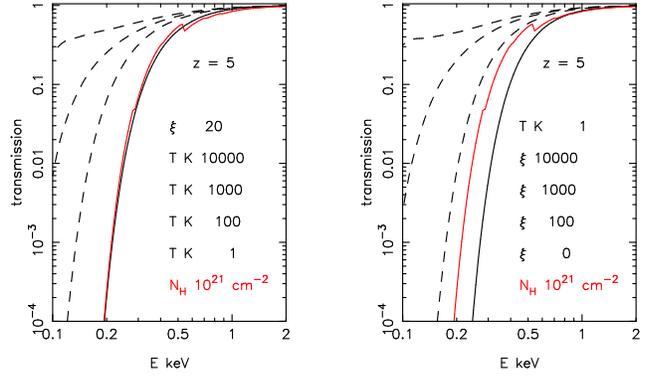}
\end{center}
\caption{Comparison of transmission for a source at $z=5$ after interaction with the IGM with metal abundance
$Z=0.2$\,Z$_{\odot}$, either varying the temperature (left panel),
or varying the ionisation parameter (right panel).
The solid red curves show the transmission through a local cold/neutral column of $10^{21}$ cm$^{-2}$ with Solar abundance, for comparison.}
\label{fig11}
\end{figure}

If the IGM is not clumpy and has the density expected to account for the
missing baryons then it is likely that it is significantly ionised.
Using Equation \ref{eq2} we can estimate
the absorption expected in a source at $z=5$ for different temperatures
and ionisation parameters. The results are shown in Fig. \ref{fig11}.
Here we have assumed a constant metallicity of $Z=0.2$\,Z$_{\odot}$ for the IGM
and no evolution of the metallicity or photoionisation environment with redshift. This metallicity is consistent with the range derived for the WHIM in simulations by Cen \& Chisari (2011) and approximately consistent with the mean WHIM metallicity of 0.1\,Z$_{\odot}$ derived in simulations by Wiersma, Schaye \& Theuns (2011). In later comparisons of this model to the GRB sample we will adopt both $Z=0.2$\,Z$_{\odot}$ and $Z=$\,Z$_{\odot}$.

Also shown for reference is the absorption from a local cold/neutral column of $10^{21}$ cm$^{-2}$. Fig. \ref{fig11} indicates that the IGM could introduce
a significant absorption signature in the soft X-ray spectra of bright high-redshift sources providing the temperature of the IGM is below $10^{5}$\,K. If the photoionisation or temperature of the IGM is higher than this the absorption is rapidly diminished.

\begin{figure}
\begin{center}
\includegraphics[height=7cm,angle=-90]{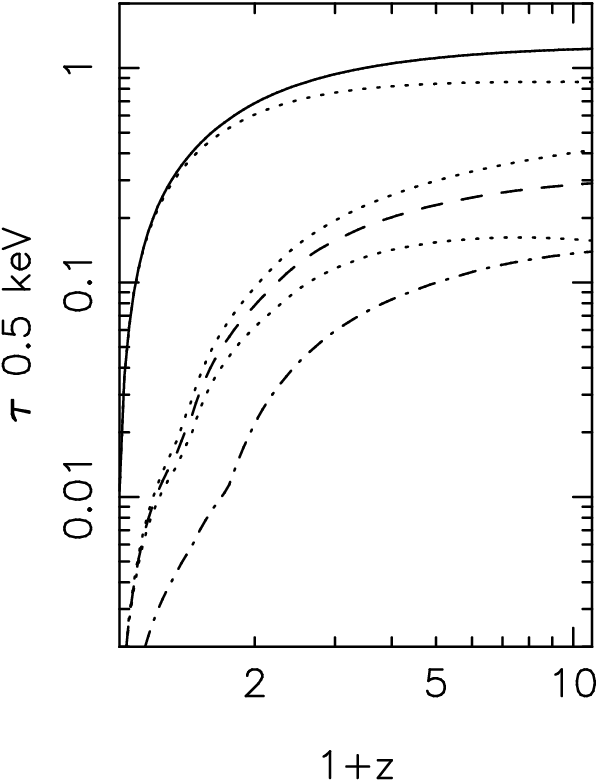}
\end{center}
\caption{The optical depth of the IGM at 0.5\,keV as a function of redshift.
A cold IGM is shown by the solid curve, while the dashed curve represents a warm IGM with $\xi=20$ erg cm s$^{-1}$
and $T=10^{4}$\,K and the dot-dashed curve represents a warm IGM
with $\xi=20$ erg cm s$^{-1}$ and $T=10^{6}$\,K - all curves have a metallicity $Z=0.2$\,Z$_{\odot}$.
The dotted curves illustrate the effects of evolution: metallicity evolving as $Z=0.2(z+1)^{-1}$ for a cold IGM (upper curve), ionisation evolving as $\xi=20(z+1)^{-1}$ (middle curve) or as $\xi=20(z+1)^{+1}$ (lower curve) for a $10^{4}$\,K IGM.}
\label{fig12}
\end{figure}

Fig. \ref{fig12} shows the increase in optical depth of the IGM at 0.5\,keV as a function of redshift. The cold curve should be compared with Fig. 4 in B11.
The current curve asymptotes to 1.2 at high-redshift rather than the value of 0.4 estimated by B11.
This is because we have 
included the hydrogen and helium cross-sections without multiplying by the metallicity and these are very significant when the medium is cold.
Even if the metallicity is set to zero the asymptote is still $\tau\approx0.8$ because of hydrogen and helium absorption alone.
The parameters used for the warm curves, $\xi=20$ erg cm s$^{-1}$, $T=10^{4}$\,K and $T=10^{6}$\,K produce the two lower curves.
In Fig. \ref{fig12} we also show the effects of metallicity and ionisation evolution.
Similar evolution profiles are obtained if the ionisation parameter is
constant but the temperature changes with redshift. We see that the bulk of the material seen by a high-redshift astrophysical source comes from redshifts of approximately 1--2. If metallicity decreases with redshift these curves asymptote to a constant value at lower redshifts.

\begin{figure*}
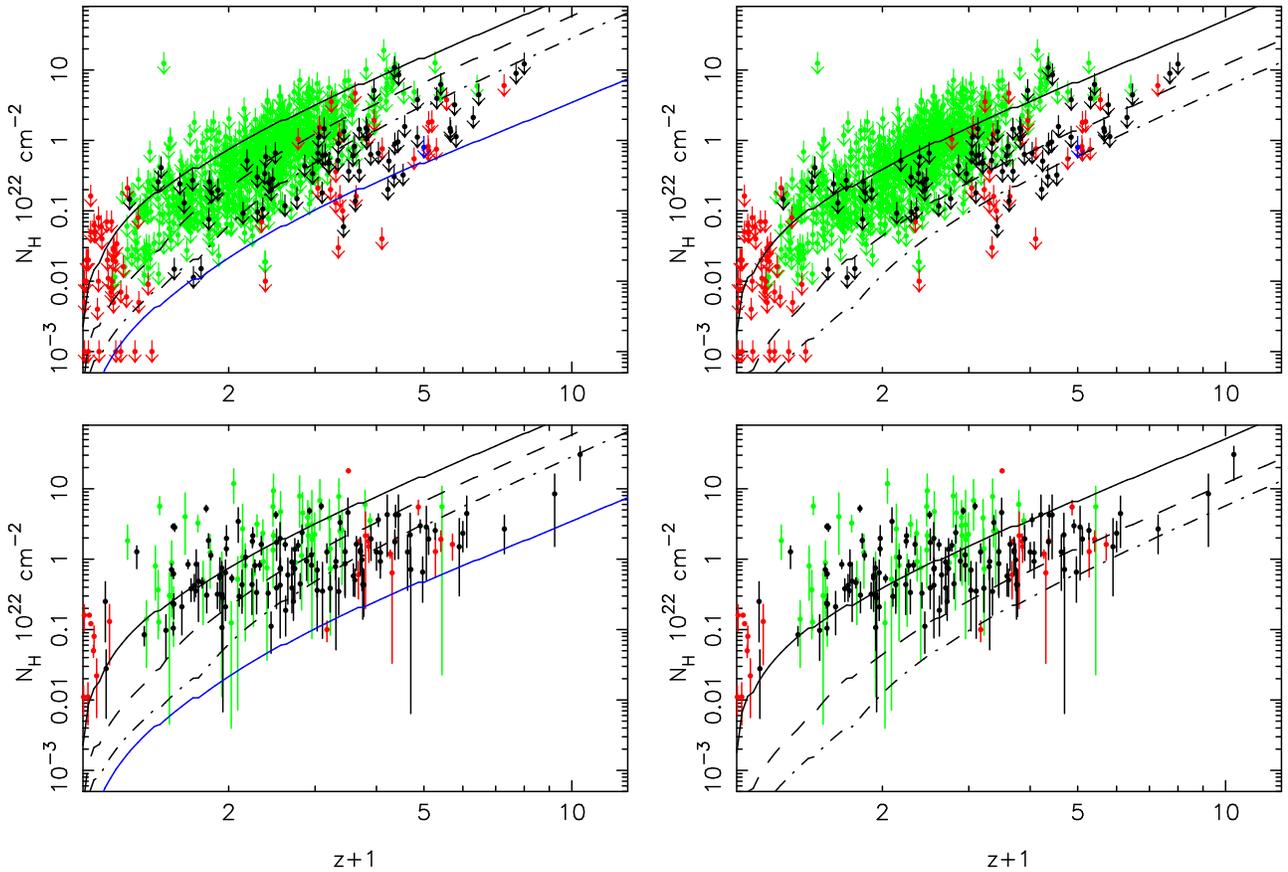

\begin{center}
\includegraphics[height=11.5cm,angle=0]{figure13a.ps}
\hspace{0.2cm}
\includegraphics[height=11.5cm,angle=0]{figure13b.ps}
\end{center}
\caption{Measured intrinsic column density at Solar metallicity for the GRBs (black points), AGN (green), quasars (red) and stacked spectrum of high-redshift quasars (blue) as described in Sections \ref{sec:nhvsz}, \ref{sec:nhvszagn} and 4.2, with constrained non-zero values in the lower panels and upper limits in the upper panels. Overlaid are the models for a cold IGM (solid line), 10$^4$\,K warm IGM (dashed line) and 10$^6$\,K warm IGM (dot-dashed line). Here we compare IGM models which use a metallicity of $Z$~=~Z$_{\odot}$ on the left and $Z$~=~0.2\,Z$_{\odot}$ on the right. 
The blue curves in the left-hand panels indicate the contribution expected from absorption by Lyman-$\alpha$ clouds, assuming a prescription for their declining metallicity with redshift as described in Section \ref{sec:lya}.}
\label{fig13}
\end{figure*}

We can convert the optical depth of the IGM at 0.5\,keV to an equivalent intrinsic hydrogen column density situated at the redshift of the host.  
Fig. \ref{fig13} shows the expected absorption curves for cold, 10$^4$\,K warm and 10$^6$\,K warm IGM, for two different metallicities. The IGM model curves are overlaid on the measured $N_{\rm H,intrinsic}-z$ relations for the GRB and AGN$+$quasar samples introduced in Sections \ref{sec:nhvsz} and \ref{sec:nhvszagn}. While the absorption observed in AGN and quasars may be of differing origin to that of the GRBs, any IGM component should contribute in the same way in all extragalactic objects, and the AGN (particularly those less absorbed, as is expected for the Type I sample of Scott et al. 2011) help constrain the IGM parameters. The cold IGM column is clearly inconsistent with many of the high-redshift objects and wholly inconsistent with all measured values at
$z>4$. However a warm IGM with a temperature in the range $10^{4}-10^{7}$\,K for $\sim$0.2\,Z$_{\odot}$ appears consistent with the observations, allowing for some scatter expected due to statistical considerations. A significant fraction of the absorption for the high-redshift objects could therefore be due to the warm IGM rather than X-ray absorbing material intrinsic to the host galaxy.

We then refitted the GRB afterglow data described in Section \ref{sec:nhvsz} with the power law model including a fixed Galactic absorption, a variable intrinsic absorption and an IGM component with fixed power law index of 1.4, density $n_{0} = 1.7 \times 10^{-7}$ cm$^{-3}$, ionisation $\xi = 20$ erg cm s$^{-1}$, Solar metallicity and temperature stepped through a wide range of temperatures from $T = 10^4 - 10^7$\,K.  
The metallicity of the intrinsic absorbing column is completely independent of that of the IGM component; we chose Solar metallicity as it is the most commonly used value despite this very likely being an overestimate, noting that intrinsic X-ray column density scales approximately with metallicity. These results indicate that an IGM temperature of 10$^7$\,K is too hot and the IGM does not absorb the X-ray flux, while IGM temperatures of $\sim10^4$\,K and below overpredict the absorption for the high-redshift sources hence they disappear from the plot as non-detections. Somewhere between 10$^5$ and 10$^{6.5}$\,K accounts for all or most of the excess X-ray column measured in the moderate to high redshift GRBs between 0.2 and $<$1\,Z$_{\odot}$ IGM (Fig. \ref{fig13}).

\subsection{Absorption by neutral gas in the IGM: Lyman-$\alpha$ absorbers} \label{sec:lya}
Absorption due to neutral hydrogen is seen in the spectra of many GRBs and quasars: 
the so called damped Lyman-$\alpha$ systems (DLA) with column densities
$>2 \times 10^{20}$ cm$^{-2}$, Lyman Limit systems (LLS) with column densities
$>10^{17}$ cm$^{-2}$ and the Lyman-$\alpha$ forest with column
densities from $10^{12}$ to $10^{17}$ cm$^{-2}$. The differential
number density per unit redshift of the clouds responsible is fit by a broken power law of the form
$f(N)=BN^{-\beta}$ with $\beta=1.83$ and $\log B=12.89$ for $N_{\rm H}<10^{16}$ cm$^{-2}$ and $\beta=1.32$ and $\log B=4.91$ when $N_{\rm H}>10^{16}$ cm$^{-2}$ (Petitjean et al. 1993). The density evolves with redshift,
$N(z)=N_{0}(1+z)^{\gamma}$ where $\gamma=2.45^{+0.75}_{-0.65}$,
(P\'eroux et al. 2003). We can integrate over column density
to calculate the total differential column per unit redshift

\begin{equation}
N_{\rm H0}=\int_{10^{12}}^{10^{22}} f(N)N_{\rm H}dN_{\rm H}
\end{equation}

and then integrate out to a given redshift to find the total column
seen in Lyman-$\alpha$

\begin{equation}
N_{\rm HLy\alpha}=\int_{0}^{z} N_{\rm H0}(1+z)^{\gamma}dz.
\end{equation}

We can compare this with the column density of the diffuse IGM considered above

\begin{equation}
N_{\rm HIGM}=
\frac{n_{0}c}{H_{0}}
\int_{0}^{z}
\frac{(1+z')^{2}}{\sqrt{(1+z')^{3}\Omega_{\rm M}+\Omega_{\Lambda}}} dz'.
\end{equation}

\begin{figure}
\begin{center}
\includegraphics[height=8.4cm,angle=-90]{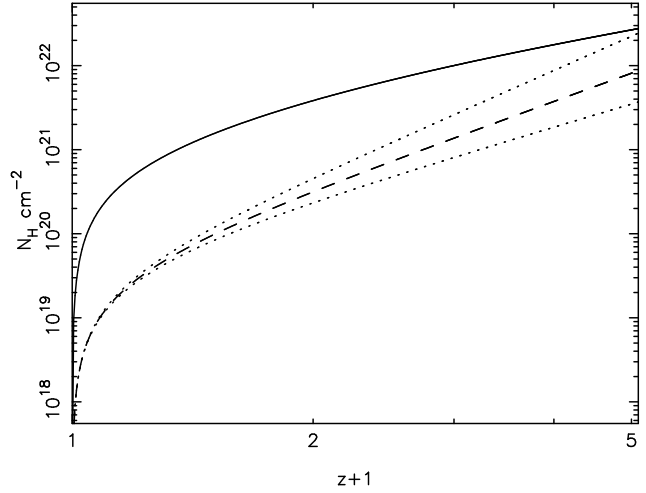}
\end{center}
\caption{The total column densities of the IGM, $N_{\rm HIGM}$ (upper solid curve) and
Lyman-$\alpha$, $N_{\rm HLy\alpha}$ (lower dashed curve). The dotted curves indicate the
uncertainty in density evolution of the Lyman-$\alpha$ absorption.}
\label{fig14}
\end{figure}

The total columns $N_{\rm HIGM}$ and $N_{\rm HLy\alpha}$ are plotted as a function of
redshift in Fig. \ref{fig14}.

At low redshift the proposed $N_{\rm HIGM}$ is a factor $\sim20$ larger
than the column seen in Lyman-$\alpha$, $N_{\rm HLy\alpha}$.
Note that the difference between $N_{\rm HIGM}$ and $N_{\rm HLy\alpha}$ diminishes as
the redshift increases because the density of cold clouds responsible for
the Lyman-$\alpha$ absorption increases with redshift.
This confirms our expectation that the bulk of the IGM column must be warm, 
$T>>13.6e/k=1.6\times10^{5}$\,K. 

In order to predict the X-ray absorption we expect from the cold Lyman-$\alpha$ clouds we need to estimate the metallicity of the clouds.
Using the metallicity evolution given by P\'eroux et al. (2007), $ \log$($Z$/Z$_{\odot}$)~=~$-0.4 z^{-0.46}$, we obtain the predicted X-ray column density versus redshift shown by the blue curve in Fig. \ref{fig13}.
This curve is not very different from that calculated for the
$T=10^6$\,K $Z$~=~0.2\,Z$_{\odot}$ WHIM model, indicating that both could be
significant components. However, it is important to emphasise that the
uncertainties in both the redshift evolution of the mass spectrum of
the intervening clouds, and their range of metallicities remain
signficant. So while the contribution from cold clouds could explain
the bulk of the extra absorption (see also Campana et al. 2012),
particularly if their average metallicities were a factor of a few higher, it
equally might turn out to be a minor factor if the average metallicity
is in fact rather lower. A final caveat is that in the case of
absorbing clouds, the column along any particular line of sight may
be dominated by a small number of individual systems, so their predicted
effect is also subject to small number statistics.

\subsection{Differences in intervening absorber properties between AGN and GRBs}
Studies of high resolution optical spectra have enabled a direct comparison between quasar intervening 
absorption systems and those in GRB sightlines. These have
suggested an excess (by a factor 2--4) of strong (restframe equivalent width $>1.0$\,\AA) intervening Mg\,{\sc II} 
absorbers in GRB afterglow spectra compared to quasar sightlines (Prochter et al. 2006; Vergani et al. 2009; Cucchiara et al. 2009b; Tejos et al. 2009). 
In contrast, no excess of C\,{\sc IV} absorbers is observed
(Sudilovsky et al. 2007; Tejos et al. 2007), but similar to Mg\,{\sc II} the number of (sub)-DLAs per unit redshift may be higher in the 
GRB afterglow sample than what is expected from quasar sightlines (Vergani et al. 2009).
The excess of intervening Mg\,{\sc II} absorbers appears to be weakly dependent on redshift (Vergani et al. 2007). Several reasons
have been put forward to explain this excess, including complex selection effects on the side of the GRBs or the quasars or both (with dust or 
lensing as candidates), but also differences in angular size of the emitting surface may play a role. This excess of absorbing systems may 
be accompanied by a noticeable associated X-ray column at the redshift of the intervening system. Campana et al. (2012) made an evaluation of the likely contribution of intervening systems to the observed X-ray column. In particular, they point out that if
the overabundance of Mg\,{\sc II} in GRB sightlines reflects the true number distribution of intervening absorbers (i.e. that the QSO sample
is affected by an observational bias that does not affect the GRB sample), but that the distribution of absorber properties (column densities) 
at all redshifts is well represented by the QSO sample, the intervening absorbers can contribute significantly to the observed X-ray column density. However, there 
is still debate about the significance of the overabundance (see Cucchiara et al. 2012 for a re-analysis of a large sample of spectra, finding a much lower
overabundance and at a lower significance level), and which sample, GRBs or QSOs, better reflects the ``true'' distribution of intervening systems along
extragalactic sightlines (see e.g. Rapoport et al.~2013).

\section{Could the absorption be dominated by hosts?}
An alternative possibility is that the IGM absorption is low and the
observed correlation between X-ray $N_{\rm H}$ and redshift in GRB afterglows represents an evolution
in GRB local environments. GRBs are created from massive stars \citep{hb11}, and lie in star forming regions within their typically
compact host galaxies \citep{fruchter06,svensson10}. Hence, evolution in the hydrogen column density to GRBs could reflect changes in the mode,
or physics of star formation over cosmic history. Some such differences can be identified. In addition to a higher universal
rate of star formation at $z\sim2$, it appears that much of the star formation arises in more massive galaxies than at the present
day, with a significant fraction occuring in a dust obscured mode \citep{chapman05,heavens04}.

Studies of individual lensed galaxies at high-redshift \citep[e.g.][]{swinbank10} suggest that in this obscured and submillimetre-bright population the star forming regions are a factor of $\sim10^7$ more luminous than in the Milky Way, although they tend to
have similar physical sizes. There are physical arguments that limit the neutral hydrogen column density that can be viewed through
a single molecular cloud, largely due to the UV field caused in denser giant molecular clouds \citep{schaye01}. This limit is
around $10^{22}$\,cm$^{-2}$, clearly below those measured in some GRBs, although it is not clear how these
effects correlate with the X-ray measured $N_{\rm H}$, which is not directly probing the neutral hydrogen. \cite{swinbank10} avoid
this concern by postulating that the giant star forming regions in high-redshift galaxies in fact contain numerous cores (to first order
$10^7$, $\sim 1$\,pc cores), which are individually rather similar to those in Milky Way. However, the overall density within such regions
means that any line of sight can intersect more than one region and hence experience a significantly higher column 
than in the same line of sight out of the Milky Way. Since these regions are compact, the UV flux from the GRB could substantially
sublimate the dust along the line of sight \citep[studies suggest the prompt UV flash may destroy dust out to $\sim 100$\,pc, ][]{fruchter01},
leading to generally low $A_V$, but still yield a higher X-ray $N_{\rm H}$. If this model is correct, and the locations of star formation
in the high-redshift universe are highly clustered, then this evolution in environment could provide a viable alternative to observing $N_{\rm H}$ within the IGM. Since the origins of AGN are less closely linked to star formation, a test of this scenario might be made if sufficient statistics on AGN and QSOs beyond $z=4$ were available.

We note that, while there may be an increasing number of
massive, dusty and high X-ray $N_{\rm H}$ hosts at redshifts $z\sim2$,
at the highest redshifts where explaining the excess columns becomes
most acute, searches for GRB host galaxies suggest they
are small (Tanvir et al. 2012b), consistent with the bulk of star-formation
at $z \gtrsim 6$ having occurred in proto-galactic fragments which had
yet to assemble into large galaxies.

\subsection{Constraints on the ionising continuum escape fraction}
Another potential important ramification of an increasing column density with redshift concerns the escape fraction 
of ionising photons. This is a critical issue for the process of reionisation, since for massive stars to be the dominant cause of
IGM ionisation seems to require that at least $\sim$20\% of the UV light produced in low-mass high-redshift galaxies escapes
the galaxies themselves (e.g. Alvarez, Finlator \& Trenti 2012).
If GRBs are produced by the same populations of stars that produce the UV then we can constrain
the escape fraction by considering the gas column densities for a sample of high-redshift bursts.
This has previously been done considering the Lyman-$\alpha$ absorbing column, which sets a minimum
level of absorption since it indicates the amount of neutral hydrogen.
Even column densities as low as $N_{\rm H}=10^{19}$\,cm$^{-2}$ will absorb a significant fraction of
the UV continuum photons.
Such studies have led to estimates of $f_{\rm esc}=0.02\pm0.02$ for 28 GRBs at $2<z<6$ (Chen, Prochaska \& Gnedin  2007);
an upper limit of $f_{\rm esc} \le 0.075$ from 33 GRBs with $z>2$ (Fynbo et al. 2009); and
$f_{\rm esc}=0.020\pm0.017$ in a composite optical spectrum of 11 GRBs at $\langle z \rangle = 3.7$ (Christensen et al. 2011).
The escape fraction can be measured for one individual object, GRB\,050908 at $z=3.35$, due to its particularly small H\,I column density, resulting in $f_{\rm esc}=0.13\pm0.04$ (Christensen et al. 2011). 

Rather than using the Lyman-$\alpha$ line, we make the assumption that the equivalent hydrogen column density as measured in the X-rays, $N_{\rm H,intrinsic}$, can provide an indication of the hydrogen column density in the immediate environment of the GRB progenitor star prior to ionisation by the GRB flux. 
This implicitly assumes that prior to the burst the gas measured by the X-ray column was largely neutral, even if it was subsequently ionised during by the
GRB itself.
Including both detections and non-detections of $N_{\rm H,intrinsic}$ in the model where all excess X-ray absorption originates in the host galaxy, we arrive at an upper limit on the mean escape fraction of $f_{\rm esc} \le 0.14$ over $0.1<z<9.4$ ($f_{\rm esc} \le 0.16$ at $z>2$) using the central values of measured $N_{\rm H,intrinsic}$. Using the lower limits instead, we derive a more conservative limit on the escape fraction of $f_{\rm esc} \le 0.39$ over $0.1<z<9.4$. 
Although the constraint we derive is not as strong as that from optical studies, essentially due to the reduced sensitivity to measuring low-columns 
via X-rays, it does make use of a larger sample, and takes account of gas which may have been ionised by the GRB and hence be missing in the optical census.
It also provides further evidence against a high UV continuum escape fraction from high-redshift star-forming galaxies. Furthermore, the apparent increase in average $N_{\rm H,intrinsic}$ with redshift, if due to evolution in the host galaxies rather than the IGM, would seem to indicate a declining rather than increasing escape fraction with redshift.

\subsection{A Population III star progenitor for $z = 8 - 9$ GRBs?} \label{sec:popiii}
The search for, and understanding of, the very first population of stars and galaxies is one of the central questions of astrophysics (e.g. Barkana \& Loeb 2001; Bromm \& Larson 2004; Ciardi \& Ferrara 2005; Glover 2005). If found, these sources will indicate the time at/over which the Universe was reionised and provide a means of studying the conditions at that crucial epoch in its history. The epoch of reionisation is currently suggested to have occurred around $z \sim 10$ as measured with the Wilkinson Microwave Anisotropy Probe (Hinshaw et al. 2012; Dunkley et al. 2009; Komatsu et al. 2009), but this is subject to very large uncertainty and debate whilst observational signatures are lacking (e.g. Persson et al. 2010). The first stars are predicted to have been very massive: $\ge 100$ M$_{\odot}$ (Population III, Abel et al. 2000; Bromm \& Larson 2004; Bromm et al. 2009) and $\ge 10$ M$_{\odot}$ (termed Population II.5 or III.2 e.g. Greif \& Bromm 2006; Tan \& McKee 2008), perhaps capable of producing a supernova and a GRB in their final demise (Heger \& Woosley 2002; Heger et al. 2003; Greif et al. 2007; M\'esz\'aros \& Rees 2010 considering $z = 20$). The immense luminosities of most GRBs, of order 10$^{51}$\,erg, mean that we are likely to be able to detect them beyond $z = 11$, perhaps to $z\sim 14-20$ (e.g. Bloom et al. 2009).

If we consider the case that all the excess X-ray column density we measure in our highest redshift GRBs, 090423 and 090429B, is intrinsic to the GRB host galaxy, then a number of arguments leads us to conclude that a Population III star progenitor for each of these GRB events seems highly unlikely, although  cannot be ruled out.
The intrinsic column densities we measure in GRBs 090423 and 090429B become very large when zero metallicity is invoked (Section \ref{sec:seds}) suggesting that metals are likely to have been present at redshifts 8--9, and therefore the environment at the epoch of these bursts was not pristine. 

Certain formation scenarios for the first stars suggest that it would be extremely difficult to obtain $\sim$10$^{24}$ cm$^{-2}$ column densities around Population III stars (e.g. Greif et al. 2009). Recent SPH simulations by Salvaterra et al. (2011) imply rapid metal enrichment occurred by $z = 7-8$. Their simulations dictate that at those redshifts, where we find GRB\,090423, galaxies can have metallicities of a tenth Solar and all long GRBs would originate in Population II stars. Our findings show that, at $Z\sim \frac{1}{10}$~Z$_{\odot}$ with no intervening absorber contribution, GRB\,090423 must have an intrinsic X-ray column density of few$\times$10$^{23}$ cm$^{-2}$, and is therefore likely to lie in a region with high local gas densities in this scenario.

Additionally, given the significant role that ionisation may play, we should consider that the X-ray column densities we measure are in fact lower limits (see e.g. Watson \& Laursen 2011). The absorbing gas close to the GRB should be strongly ionised by the blast of high energy radiation from the GRB jet, while what we measure is the equivalent neutral hydrogen column density.  
Crude lower limits on the metallicity can be made by assuming the material is Thomson thin (e.g. Campana et al. 2011) and all located within the GRB host galaxy. Using this method on GRBs 090423 and 090429B we find $Z>0.05$ and $Z>0.03$ respectively (using the 90\% uncertainties on $N_{\rm H,intrinsic}$ from the best-fitting models to the SEDs). 

In summary, should all the observed X-ray absorption originate in the circumburst medium, a Population III progenitor for the GRBs in our sample at redshift 8 and above is unlikely given the large column densities this would imply. The $N_{\rm H} - z$ relation we investigate here implies that at even higher redshifts where we should find the first stars, column densities will remain higher than those achieved in current Population III star scenarios when zero metallicity is invoked. A combination of intrinsic and intervening absorption contributing to the spectral shape observed in these high-redshift sources may therefore be preferable.

\section{Conclusions} \label{sec:discuss}
We have revisited the puzzle of the apparent increase in $N_{\rm H,intrinsic}$ with redshift
seen in GRB afterglow X-ray spectra. We have reanalysed a large sample of GRBs with redshifts,
paying particular attention to the highest redshift bursts, and find the correlation to exist 
and not to be explicable by systematic biases or selection effects. The trend may also be seen in a sample of AGN$+$quasars.

We have outlined a model for a diffuse, warm-hot IGM and find this can account for most or all of the measured X-ray absorbing column density in $z \ge 3$ GRBs, providing it was relatively metal enriched ($Z\sim$~Z$_{\odot}/5$), and not too hot ($T\sim10^5-10^6$\,K).
Our model, whilst still simplified, is a significant improvement on the cold, neutral IGM considered by B11. An estimate of the absorption due to discrete Lyman-$\alpha$ gas clouds suggests
they could also provide a comparable fraction of the apparent X-ray $N_{\rm H}$
in high-redshift GRB spectra, although the relative importance of these two IGM 
components remains uncertain due to our incomplete knowledge of the metal
enrichment of either.

If it is correct that the WHIM dominates the measured $N_{\rm H,intrinsic}$ for moderate to high redshift GRBs, then it will be one of the first observational windows on this elusive component which is likely to 
comprise a major proportion of the baryon budget of the Universe.
There is a well documented discrepancy between neutral hydrogen columns as measured directly from optical data and the equivalent hydrogen columns implied by measurements in the X-ray band (e.g. Galama \& Wijers 2001; Stratta et al. 2004; Kann et al. 2006; Starling et al. 2007; Schady et al. 2010). While this could be due to ionisation by the GRB since the X-ray band predominantly measures metals (e.g. Watson et al. 2007; Schady et al. 2011), this could also be at least partially alleviated with inclusion of the IGM component. 

If, instead, we consider the observed excess X-ray absorption is intrinsic to the host galaxies, we find that for the highest redshift GRBs these columns become very large, and only increase as more realistic, lower metallicities are used. This implies that a larger reservoir of material surrounds the progenitor at higher redshifts. Comparison of our model fits at zero metallicity to GRBs 090423 and 090429B with simulations of Population III stars indicates that we need to look beyond $z=9$ to witness the deaths of the first stars as GRBs, if such events exist, since $Z$~=~0 pushes the intrinsic column density to unrealistically large values. 

Our `best bet' scenario is, therefore, one in which absorption
by a WHIM component begins to be a significant
contributor to the measured X-ray $N_{\rm H}$ at $z\gtrsim 3$, and is usually
dominant at $z \sim 6$. This absorption
could be due either to a WHIM with the characteristics described and/or
to the cumulative effect of cold Lyman-$\alpha$ gas clouds on the line
of sight. Without such an integrated component, it becomes very hard to
explain the extremely high columns measured for the highest
redshift GRBs, given that they reside in small, presumably low-metallicity, galaxies.
At lower redshifts ($z \lesssim 3$) the lower envelope provided
by the WHIM absorption is significantly less than the typical measured X-ray $N_{\rm H}$ for GRBs,
and so it is likely here that absorption within the host (both proximate
to the GRB and at greater distances) dominates. Thus any correlation
between $N_{\rm H}$ and redshift in this regime must reflect changes
in the typical host properties (size, density and metallicity, for example)
and/or sample selection effects.

The ultimate tests of the IGM theory would be high-resolution X-ray spectroscopy that could detect absorption edges from individual ions, and/or definitive evidence for variable column density as may be expected soon after the GRB event due to ionisation of the surrounding medium. Unfortunately, neither are yet possible. The latter has been claimed for a small number of GRBs, but other spectral evolution effects with similar manifestations can be present as shown in Butler \& Kocevski (2007). Alternatively, if even a single high-redshift source were found with a very strong $N_{\rm H}$ upper limit, well below the lower envelope expected from smooth warm-IGM absorption, then it would effectively rule out that model as an explanation for most of the $N_{\rm H}-z$ correlation.

Proposed X-ray missions like {\em ORIGIN} (den Herder et al. 2012) are designed to make high spectral resolution measurements of the soft X-ray spectra of GRB
afterglows in the hope of detecting absorption edge features (See also Branchini et al. 2009). Such observations would be able to tell us the distribution of the material along the line of sight and the physical state of the plasma, its temperature, composition and density.
If the material is distributed in filaments rather than evenly along the line of sight as considered here these measurements could observe
the X-ray equivalent of the Lyman-$\alpha$ forest in the warm-hot material that occupies the voids in between the dense neutral hydrogen clouds.

The relationship between $N_{\rm H}$ and $z$ is highly dependent on the few $z>5$ GRBs currently at hand. 
The detection of very high redshift GRBs is proving to be of great value, and while we have not yet reached the earliest epoch of the formation of stars GRBs arguably show the greatest promise for doing so. Studying bursts at the highest redshifts is crucially dependent upon the NIR rapid response instrumentation to detect the reddened afterglows, and IR spectrographs to detail the chemical composition of the host galaxies. At present these opportunities are limited by the small number of IR spectrographs mounted on large aperture telescopes (Manchado 2003 provides an informative census of IR instrumentation during the early part of the last decade). Proposals for future space-based missions, one example being {\em Lobster}, have included the provision of IR telescopes alongside X-ray instruments, capable of identifying the afterglows of high redshift GRBs. Even if one of these is launched, ground-based follow-up is the next link in the chain to deeply probe the host galaxies and identify the progenitor stars of these most distant bursts.

\section{Acknowledgments}
We thank Andy Beardmore, Phil Evans, Lucy Heil and Kim Page for useful discussions, and the anonymous referee for helpful comments. 
RLCS is supported by a Royal Society Dorothy Hodgkin Fellowship. AES and KW acknowledge support from STFC. We gratefully acknowledge funding for {\em Swift} at the University of Leicester from the UK Space Agency. 
This work made use of data supplied by the UK {\it Swift} Science Data Centre at the University of Leicester.

\begin{table*}
\caption{~Results (main parameters) of fits to the SEDs for GRB\,090423 at $z=8.23$ and GRB\,090429B at $z=9.4$. We also list fits to the {\it Swift} XRT X-ray spectrum only for comparison.
Metallicity is given in brackets after the model name: 1 = Solar, 0.33 = LMC, 0.125 = SMC. All errors are quoted at the 90 per cent confidence level (1.6$\sigma$). The F-test probability in column 8 gives the probability that the result is obtained by chance, therefore a significant improvement in the fit when adding one extra free parameter is indicated by a low probability. U denotes an unconstrained parameter with central fitted value in brackets. $^*$ indicates the inclusion of extrapolated $K$-band data. $^+$ indicates the use of different atomic data to enable direct comparison with other works.}
\begin{center}
\begin{tabular}{l l l l l l l l}
Model($Z$/Z$_{\odot}$) & $N_{\rm H, int}$& $E(B-V)_{\rm int}$&$\Gamma_1$& $\Gamma_2$ & $E_{\rm bk}$&$\chi^2$/dof& F-test probability\\
 & ($\times$10$^{22}$ cm$^{-2}$) & (mag) & & & (keV) & \\
\hline\hline
GRB\,090423: &&&&&&& \\ \hline
 pl+smc(1)&$\le 3$&$\le 0.005$&1.58$\pm0.01$&-&- &65.9/32&\\
 pl+smc(1)$^*$&$\le 3$&$\le 0.002$&1.57$\pm0.01$&-&- &74.7/33&\\
 bknpl+smc(1)&16$^{+11}_{-9}$&$\le 0.22$& 1.26$^{+0.21}_{-1.24}$&1.97$^{+0.16}_{-0.15}$&0.06$^{+1.68}_{-0.06}$ &38.0/30&10$^{-4}$ (cf pl+smc(1))\\
 bknpl+smc(1)$^*$&16$^{+11}_{-9}$&$\le 0.06$&1.3$^{+0.1}_{-0.7}$&1.97$^{+0.16}_{-0.15}$&0.06$^{+0.15}_{-0.06}$&38.7/31&\\
 bknpl+smc(1)&16$^{+11}_{-9}$&0 fixed& 1.26$^{+0.29}_{-0.21}$&1.97$^{+0.16}_{-0.15}$&0.06$^{+1.69}_{-0.05}$ &38.7/31&\\
 bknpl+smc(1)$^*$&16$^{+11}_{-9}$&0 fixed& 1.26$^{+0.13}_{-0.14}$&1.97$^{+0.16}_{-0.15}$&0.07$^{+1.3}_{-0.05}$ &38.7/32&\\
 bknpl+smc(1)$^+$&10$^{+7}_{-6}$&0 fixed& 1.26$^{+0.13}_{-0.14}$&1.99$^{+0.16}_{-0.15}$&0.07$^{+1.30}_{-0.05}$ &38.0/31&\\
 bknpl+smc(1)&12$^{+9}_{-7}$&$\le 0.008$&= $\Gamma_2$ - 0.5 &1.90$^{+0.16}_{-0.12}$&0.13$^{+1.71}_{-0.11}$ &40.4/31&\\
 bknpl+smc(1)$^*$ & 10$^{+8}_{-6}$ & $\le 0.006$ & = $\Gamma_2$ - 0.5& 1.85$^{+0.10}_{-0.09}$&0.07$^{+0.26}_{-0.06}$ &41.5/32\\
 bknpl+smc(1)&13$\pm8$&0 fixed&= $\Gamma_2$ - 0.5 &1.90$^{+0.16}_{-0.12}$&0.13$^{+1.71}_{-0.11}$ &40.4/32&\\
 bknpl+smc(1)$^*$ & 10$^{+8}_{-6}$ & 0 fixed & = $\Gamma_2$ - 0.5& 1.85$^{+0.10}_{-0.09}$&0.07$^{+0.26}_{-0.05}$ &41.5/33\\
 bknpl+smc(0.33)&39$^{+28}_{-21}$&0 fixed&1.26$^{+0.29}_{-0.21}$&1.96$\pm0.15$&0.06$^{+1.68}_{-0.04}$ &38.6/31&\\
 bknpl+smc(0.125)&71$^{+50}_{-39}$&0 fixed&1.26$^{+0.29}_{-0.21}$&1.94$^{+0.15}_{-0.13}$&0.06$^{+1.68}_{-0.04}$ &38.5/31&\\
 bknpl+smc(0.01)&127$^{+91}_{-69}$&0 fixed&1.26$^{+0.29}_{-0.21}$&1.91$^{+0.14}_{-0.12}$&0.05$^{+1.73}_{-0.03}$ &38.4/31&\\
 bknpl+smc(0)&137$^{+97}_{-75}$&0 fixed&1.26$^{+0.29}_{-0.21}$&1.91$^{+0.13}_{-0.12}$&0.04$^{+1.74}_{-0.03}$ &38.4/31&\\ \hline

X-ray only:& & & & & & & \\
XRT pl (1) & 16$^{+11}_{-9}$ & - & - & 1.97$^{+0.16}_{-0.15}$ & - & 38.7/31 & \\
XRT pl (1)$^+$ & 10$^{+7}_{-5}$ & - & - & 1.99$^{+0.16}_{-0.15}$ & - & 39.0/31 & \\\hline \hline

GRB\,090429B: &&&&&&& \\ \hline
pl+smc(1)&24$^{+16}_{-11}$&0.05$^{+0.02}_{-0.03}$&1.83$\pm$0.03&-&- &12.0/16&\\
pl+mw(1)&24$^{+15}_{-12}$&0.08$^{+1.10}_{-0.04}$&1.81$^{+0.03}_{-0.02}$&-&- &12.1/16&\\
pl+smc(1)$^+$&15$^{+9}_{-7}$&0.05$^{+0.02}_{-0.03}$&1.83$\pm$0.03&-&- &11.9/16&\\
bknpl+smc(1)&28$^{+12}_{-13}$&0.05$^{+0.02}_{-0.03}$&1.82$\pm$0.03&U(4.5)&U(9) &9.9/14&0.26 (cf pl+smc(1)) \\
pl+smc(0.33)&66$^{+41}_{-32}$&0.05$^{+0.02}_{-0.03}$&1.83$\pm$0.03&-&- &12.0/16&\\
pl+smc(0.125)&131$^{+88}_{-65}$&0.05$^{+0.02}_{-0.03}$&1.83$\pm$0.03&-&- &12.0/16&\\
pl+smc(0.01)&288$^{+237}_{-151}$&0.05$^{+0.02}_{-0.03}$&1.83$\pm$0.03&-&- &12.1/16&\\
pl+smc(0)&319$^{+275}_{-169}$&0.05$^{+0.02}_{-0.03}$&1.83$\pm$0.03&-&- &12.3/16&\\ \hline
X-ray only:& & & & & & & \\
XRT pl (1) & 36$^{+41}_{-27}$ & - & - & 1.96$^{+0.37}_{-0.30}$ & - & 10.8/14 & \\ 
XRT pl (1)$^+$ & 22$^{+26}_{-17}$ & - & - & 1.97$^{+0.37}_{-0.30}$ & - & 10.7/14 & \\ \hline \hline
\end{tabular} 
\end{center}
\label{tab:SEDfits}
\end{table*}

%\bsp

\label{lastpage}

\end{document}